\UseRawInputEncoding

\documentclass[twocolumn]{aastex63}
\usepackage{graphicx}
\usepackage{amsmath}
\usepackage{cases}
\usepackage{color}

\defcitealias{Minami:2020a}{MK20}

\DeclareMathAlphabet{\mathsc}{OT1}{cmr}{m}{sc}
\def\testbx{bx}%
\DeclareRobustCommand{\ion}[2]{%
\relax\ifmmode
\ifx\testbx\f@series
{\mathbf{#1\,\mathsc{#2}}}\else
{\mathrm{#1\,\mathsc{#2}}}\fi
\else\textup{#1\,{\mdseries\textsc{#2}}}%
\fi}

\newcommand{\beq}{\begin{equation}}
\newcommand{\eeq}{\end{equation}}

\newcommand{\hi}{H{\sc i}~}
\newcommand{\HI}{H{\sc i}}

\newcommand{\nhi}{${N}_\mathrm{H{\sc I}}~$}
\newcommand{\nHI}{${N}_\mathrm{H{\sc I}}$}

\newcommand{\THI}{$T_\mathrm{H\textsc{i}}$}
\newcommand{\Bhi}{$B_\mathrm{H\textsc{i}}~$}
\newcommand{\BHI}{$B_\mathrm{H\textsc{i}}$}
\newcommand{\Ehi}{$E_\mathrm{H\textsc{i}}~$}
\newcommand{\EHI}{$E_\mathrm{H\textsc{i}}$}
\newcommand{\Qhi}{$Q_\mathrm{H\textsc{i}}~$}
\newcommand{\QHI}{$Q_\mathrm{H\textsc{i}}$}
\newcommand{\Uhi}{$U_\mathrm{H\textsc{i}}~$}
\newcommand{\UHI}{$U_\mathrm{H\textsc{i}}$}

\newcommand{\thetahi}{$\theta_\mathrm{H{\sc I}}$~}
\newcommand{\thetaHI}{$\theta_\mathrm{H{\sc I}}$}
\newcommand{\Deltatheta}{$\Delta\theta(\mathrm{H\textsc{i}},353)$}
\newcommand{\DeltathetaS}{$\Delta\theta(\mathrm{H\textsc{i}},353)$~}

\begin{document}

\title{\large The Origin of Parity Violation in Polarized Dust Emission\\ and Implications for Cosmic Birefringence}

\author[0000-0002-7633-3376]{S. E. Clark}
\affiliation{Institute for Advanced Study, 1 Einstein Drive, Princeton, NJ 08540, USA}
\email{seclark@ias.edu}

\author[0000-0003-2896-3725]{Chang-Goo Kim}
\affiliation{Department of Astrophysical Sciences, Princeton University, 4 Ivy Lane, Princeton, NJ 08544, USA}

\author[0000-0002-9539-0835]{J.~Colin Hill}
\affiliation{Department of Physics, Columbia University, New York, NY, USA 10027}
\affiliation{Center for Computational Astrophysics, Flatiron Institute, New York, NY, USA 10010}

\author[0000-0001-7449-4638]{Brandon S. Hensley}
\affiliation{Department of Astrophysical Sciences, Princeton University, 4 Ivy Lane, Princeton, NJ 08544, USA}
\affiliation{Spitzer Fellow}

\begin{abstract}
Recent measurements of Galactic polarized dust emission have found a nonzero $TB$ signal, a correlation between the total intensity and the $B$-mode polarization component. 
We present evidence that this parity-odd signal is driven by the relative geometry of the magnetic field and the filamentary interstellar medium in projection.  
Using neutral hydrogen morphology and {Planck} polarization data, we find that the angle between intensity structures and the plane-of-sky magnetic field orientation is predictive of the signs of Galactic $TB$ and $EB$. 
Our results suggest that magnetically misaligned filamentary dust structures introduce nonzero $TB$ and $EB$ correlations in the dust polarization, and that the intrinsic dust $EB$ can be predicted from measurements of dust $TB$ and $TE$ over the same sky mask. We predict correlations between $TE$, $TB$, $EB$, and $EE/BB$, and confirm our predictions using synthetic dust polarization maps from magnetohydrodynamic simulations. We introduce and measure a scale-dependent effective magnetic misalignment angle, $\psi_\ell^{dust} \sim 5^\circ$ for $100 \lesssim \ell \lesssim 500$, and predict a positive intrinsic dust $EB$ with amplitude $\left<D_\ell^{EB}\right> \lesssim 2.5~\mu\mathrm{K^2_{CMB}}$ for the same multipole range at 353\,GHz over our sky mask. Both the sign and amplitude of the Galactic $EB$ signal can change with the sky area considered. 
Our results imply that searches for parity violation in the cosmic microwave background must account for the nonzero Galactic $EB$ and $TB$ signals, necessitating revision of existing analyses of the evidence for cosmic birefringence. 
\end{abstract}

\section{Introduction}

The polarized sky at microwave frequencies consists, at minimum, of radiation from the cosmic microwave background (CMB) and dust and synchrotron emission from the Milky Way. On the celestial sphere, the observed Stokes $Q$ and $U$ parameters describing the linear polarization field can be decomposed into two rotationally invariant quantities that behave differently under a parity transformation: an $E$-mode component that does not change sign, and a $B$-mode component that does. This decomposition is motivated by the study of the polarized CMB, because scalar perturbations in the early universe generate only $E$-mode fluctuations at linear order, while tensor perturbations --- a prediction of inflationary cosmology --- generate both $E$- and $B$-mode fluctuations at the surface of last scattering \citep{Kamionkowski:1997a, Kamionkowski:1997b, Seljak:1997}.

The primordial $B$-mode polarization signal has not yet been detected, and is known to be subdominant at all frequencies to polarized Galactic emission across the full sky \citep[e.g.,][]{Flauger:2014,BICEP2/KeckCollaboration:2015,PlanckCollaborationXXX:2016,2018PhRvL.121v1301B,PlanckCollaborationXI:2020}. At frequencies~$\gtrsim 100$ GHz, the Galactic polarization is dominated by dust emission: partially polarized thermal emission from interstellar dust grains that are preferentially aligned with their short axes parallel to the ambient magnetic field \citep{Purcell:1975}. The $E$- and $B$-mode polarization from Galactic dust emission thus probe the magnetic interstellar medium (ISM). Characterizing this emission is important for understanding the interplay between matter and magnetic fields in the ISM, as well as for foreground mitigation for CMB experiments. 
 
Statistical quantities of interest include the cross- or auto-power spectra of the polarized emission: $C_\ell^{XY}$, which we will refer to with the shorthand $XY$, where $X$ and $Y$ are any of $T$ (total intensity), $E$, and $B$. The {Planck} satellite mapped the whole sky in nine frequency bands, including seven that were sensitive to polarization \citep{PlanckCollaborationI:2020}. These maps enable measurements of the polarized cross-power spectra of Galactic emission, particularly at 353 GHz, the highest-frequency polarization-sensitive {Planck} channel and the channel most sensitive to polarized dust emission. 
For the diffuse sky at 353 GHz the {Planck} data at large angular scales exhibit several statistical properties of note: an overall asymmetry in the amplitude of $E$- and $B$-mode power in the Galactic emission ($EE/BB \sim 2$), a positive cross-correlation between the total intensity and the $E$-mode polarization ($TE > 0$), and a weakly positive $TB$, the cross-correlation between total intensity and the $B$-mode polarization \citep{PlanckCollaborationXXX:2016, PlanckCollaborationXI:2020}. $EE$, $BB$, and $TE$ are invariant under a parity transformation, but this property is not shared by $TB$, the correlation between the scalar intensity and the parity-odd component of the polarization. $EB$ is also a parity-odd quantity, but is consistent with null in the {Planck} data, within the statistical errors \citep{PlanckCollaborationXI:2020}.

What is the physical origin of these statistical correlations? The possible relationship between these correlations and the turbulent properties of the ISM is an area of active study \citep{Caldwell:2017, Kandel:2017, Kritsuk:2018, Kim:2019}. Both the nonunity $EE/BB$ and positive $TE$ correlations are thought to originate, at least on some angular scales, from the preferential alignment between anisotropic density structures and the interstellar magnetic field \citep{Clark:2015, PlanckXXXVIII:2016}. 
This interpretation is strongly supported by investigations based on the structure of 21-cm neutral hydrogen (\HI) emission. This line of inquiry began with the discovery that slender linear features in high-resolution \hi maps are extremely well aligned with the ambient magnetic field as traced by starlight polarization \citep{McClure-Griffiths:2006, Clark:2014} and polarized thermal dust emission \citep{Clark:2015, Martin:2015}. Indeed, template maps constructed solely from \hi orientation and dust total intensity can reproduce the $EE/BB$ asymmetry \citep{Clark:2015}. The geometry of \hi emission alone is predictive of a number of statistical properties of dust polarization, including the $EE/BB$ ratio and positive $TE$ correlation \citep{ClarkHensley:2019}.

Similar phenomenology has not yet been observationally linked to the nonvanishing parity-odd $TB$ correlation. 
While there is no \textit{a priori} reason that the observed Galactic polarization must be parity invariant,\footnote{Here we refer to the parity properties of the observed sky, which can be a particular realization of a parity-invariant underlying theory.} there is also no well-motivated physical model that predicted this parity violation, nor its observed scale dependence. 
As the ISM is sculpted by magnetohydrodynamic (MHD) turbulence, the $TB$ signal could plausibly be related to some parity-odd MHD quantity, e.g., the magnetic helicity or cross-helicity \citep{Brandenburg:2005, Blackman:2015}.
Toy models of a large-scale helical magnetic field can produce nonzero $TB$ and $TE$ signals at very low multipoles \citep[$\ell < 22$;][]{Bracco:2019Helical}, but the observed $TB$ and $TE$ spectra are much flatter than predicted by these models \citep{Huffenberger:2020}. 

The empirical relationship between magnetically aligned density structures and the $TE$ and $EE/BB$ correlations motivates consideration of a filament-based explanation for nonzero $TB$. Idealized filaments with polarized emission that is either parallel or perpendicular to the long axis of the filament will produce predominantly $E$-like polarization, while a $45^\circ$ angle between the filament axis and the polarization angle preferentially generates $B$-like polarization \citep{Zaldarriaga:2001, Rotti:2019, Huffenberger:2020}. If filamentary dust emission is the correct model for production of both nonzero $TE$ and nonzero $TB$, it implies a nonzero Galactic $EB$ signal as well. 

Measuring a nonzero $EB$ correlation in the primary CMB would be evidence for parity-violating physics beyond the standard model of cosmology, such as cosmic birefringence \citep[e.g.,][]{1998PhRvL..81.3067C}, or of non-trivial symmetry-breaking properties in the physics of inflation~\citep[e.g.,][]{2009PhRvL.102s1302W,2011MNRAS.412L..83W,2010PhRvD..81j3532D}. Imperfect calibration of the overall angle of a polarimeter will also generate nonzero $EB$; a frequent practice is to correct for this systematic error by forcing $EB$ to vanish (at CMB-dominated frequencies), under the hypothesis that the primordial $EB = 0$~\citep{2010PhRvD..81f3512Y,2013ApJ...762L..23K,Abitbol:2016}. This process, known as ``self-calibration," removes any sensitivity to an overall cosmic birefringence angle in the data, although the power spectrum of the birefringence fluctuations can still be constrained~\citep{2014PhRvD..89f2006K,2015PhRvD..92l3509A,2017PhRvD..96j2003B,Namikawa:2020,2020PhRvD.102h3504B,2020JCAP...11..066G}. Alternatively, by using instrument modeling and/or {\it in situ} measurements, one can calibrate the polarimeter angle independently and thus constrain an overall cosmic birefringence angle via the observed $EB$ and $TB$. The latest constraints from {Planck}~\citep{2016A&A...596A.110P} and the Atacama Cosmology Telescope using this method find results consistent with null \citep{Choi:2020}.

The possibility of a nonzero Galactic $EB$ signal further complicates attempts to measure a primordial $EB$ correlation, as well as the self-calibration technique that assumes the intrinsic CMB $EB$ and $TB$ signals vanish \citep{Abitbol:2016}. \citet{Minami:2019} introduced a formalism for simultaneous determination of the instrument miscalibration and cosmic birefringence angles, taking advantage of the fact that the Galactic foreground polarization is rotated by the instrument miscalibration angle only, while the CMB polarization is rotated by both the miscalibration and the cosmic birefringence. \citet{Minami:2020a} (hereafter \citetalias{Minami:2020a}) used this methodology to find evidence for an isotropic cosmic birefringence angle $\beta$ in {Planck} data at $2.4\sigma$ significance ($\beta = 0.35 \pm 0.14$), with the additional assumption that the Galactic $EB$ = 0. If the Galactic $EB$ is nonzero, the evidence for cosmic birefringence must be reevaluated in light of this foreground signal. 

In this paper, we use information derived from \hi data to present evidence that imperfect alignment between filamentary dust structures and the sky-projected magnetic field is the origin of nonzero $TB$ and $EB$ in the Galactic dust emission. In Section~\ref{sec:data} we introduce the data used in this analysis. In Section~\ref{sec:globallynonzero} we demonstrate that the measurement of nonzero Galactic $TB$ over the diffuse sky is robust. In Section~\ref{sec:origin}, we describe our hypothesis for the origin of parity-odd quantities in dust polarization (\ref{sec:hypothesis}) and demonstrate support for our hypothesis in {Planck} data for the $TB$ signal (\ref{sec:TBdeltatheta}) and for the $EB$ signal (\ref{sec:EBdeltatheta}). We interpret these results in Section \ref{sec:interpretation} and test further predictions of our model in both {Planck} data and MHD simulations in Section \ref{sec:correlation}. 
We discuss the implications of our results for cosmic birefringence searches in Section~\ref{sec:cosmicbirefringence} and conclude in Section~\ref{sec:conclusions}. 

\begin{figure*}
    \centering
\includegraphics[width=\textwidth]{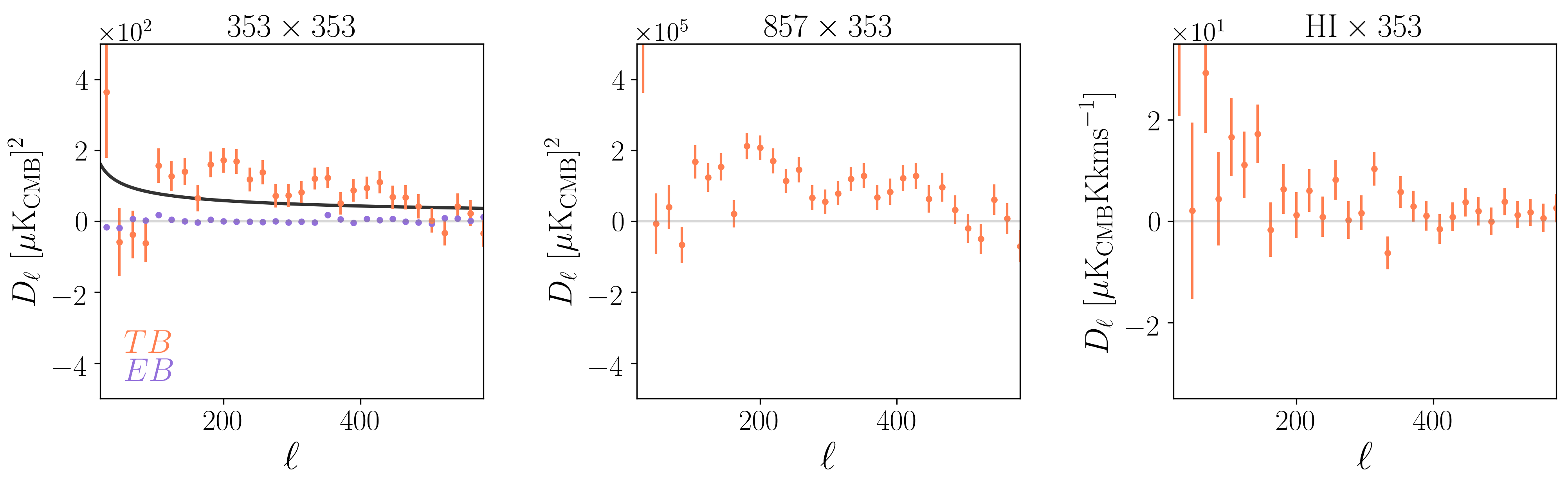}
\caption{$TB$ correlations (orange) computed with $T$ total intensity maps, left to right: \texttt{NPIPE} 353 GHz ($T_{353}$), \texttt{NPIPE} 857 GHz ($T_{857}$), neutral hydrogen intensity (\THI). $B_{353}$ is computed from the \texttt{NPIPE} 353 GHz A/B splits. Leftmost panel also shows $E_{353}B_{353}$ computed from \texttt{NPIPE} A/B splits (purple). Black line in leftmost panel shows the fit to $T_{353}B_{353}$ calculated for PR3 data in \citet{PlanckCollaborationXI:2020}. }
\label{fig:TB_353_857_HI}
\end{figure*}

\section{Data}\label{sec:data}

\subsection{Planck data}
We use several publicly available data products produced by the {Planck} collaboration. The analysis in this work uses the recently released maps produced using the \texttt{NPIPE} processing pipeline \citep{PlanckCollaborationI:2020}. We use observations taken with the {Planck} High Frequency Instrument \citep{PlanckCollaborationIII:2020}. We principally make use of the \texttt{NPIPE} 857, 353, and 217\,GHz A/B data splits, which were independently observed with different horns and are expected to have largely uncorrelated systematics. The \texttt{NPIPE} maps are all released in $\mathrm{K}_\mathrm{CMB}$ temperature units, and the conversion factor to flux density units for the 857 GHz maps is 2.27 MJy sr$^{-1} / \mathrm{K}_\mathrm{CMB}$ \citep{PlanckCollaborationIX:2014, PlanckCollaborationLVII:2020}. We further confirm our results using the {Planck} Data Release 3 (PR3) maps \citep{PlanckCollaborationI:2020}, to ensure that our conclusions are not qualitatively affected by the choice of {Planck} processing pipeline. 

\subsection{{\rm \HI}-based Stokes parameter maps }
Observations of \hi are a valuable tool for deciphering the physical nature of dust emission. 
The column densities of \hi and dust are linearly correlated in the diffuse ISM \citep{Boulanger:1996, Lenz:2017}, and thus the column density of \hi is a useful proxy for the dust column. 
Recent work has shown a deeper link between these two tracers: the morphology of \hi structure probes the polarization structure of the dust emission. Linear structures in diffuse \hi emission are well aligned with the plane-of-sky magnetic field orientation traced by the dust polarization angle \citep{Clark:2015, Martin:2015}. The magnetic alignment is driven by density structures with properties consistent with the cold neutral medium phase of \hi \citep{McClure-Griffiths:2006, Clark:2019, PeekClark:2019, Kalberla:2020, Murray:2020}. 
Broadband measurements of dust polarization measure the line-of-sight integrated dust emission projected onto the plane of the sky, and thus changes in the polarization angle along the line of sight within a telescope beam contribute to depolarization of the measured dust signal. \citet{Clark:2018} showed that this line-of-sight magnetic field tangling can be predicted from the coherence of \hi orientation as a function of \hi velocity. \citet{Pelgrims:2021} used this property of \hi in conjunction with a Gaussian decomposition of \hi data \citep{Panopoulou:2020} to detect line-of-sight frequency decorrelation in {Planck} data. 

\citet{ClarkHensley:2019} used these empirical relationships between \hi and dust to construct three-dimensional (position-position-velocity) maps of the Stokes parameters of linear polarization predicted solely from the morphology of \hi emission. These maps are integrated over line-of-sight velocity to produce \Qhi and \UHI, \HI-based sky maps of the Stokes parameters of linear polarization. From these, the plane-of-sky magnetic field orientation inferred from the \hi geometry is \thetahi = $\frac{1}{2}\mathrm{arctan}\frac{U_\mathrm{H\textsc{i}}}{Q_\mathrm{H\textsc{i}}}$. \citet{ClarkHensley:2019} compute two sets of maps using different \hi surveys; in this analysis we use the $16.2'$ all-sky maps constructed from the \HI4PI survey \citep{HI4PICollaboration:2016}. The \Qhi and \Uhi maps are integrated over $-90 < v_{lsr} < 90$ km/s. We also make use of \THI, the \hi total intensity over this same velocity range.  

\subsection{Sky masks}\label{sec:skymasks}

The primary results presented in this work are for cross-power spectra computed on the {Planck} $70\%$ sky fraction Galactic plane mask \citep{PlanckCollaborationXIX:2015}. We apodize this mask with a $60'$ cosine taper such that our final mask has $f_{sky} = \frac{1}{N} \sum_i^{N} w_i^2 \sim 0.69$, where $N$ is the number of map pixels and $w_i$ is the fractional weight of each pixel.

The results demonstrating the origin of $TB$ and $EB$ in Galactic dust emission (e.g., Sections \ref{sec:TBdeltatheta} and \ref{sec:EBdeltatheta}) are qualitatively unchanged for a simple sky mask defined by $|b| > 30^\circ$, and are similarly insensitive to the additional application of the {Planck} 353 GHz polarization point source mask \citep{PlanckCollaborationIX:2014}. However, one of the important results of our work is that parity-odd quantities in dust polarization differ depending on the sky area considered. We note that our fiducial $f_{sky}\sim 0.69$ sky mask is very different from the sky masks considered in \citetalias{Minami:2020a}. The ramifications of this difference are discussed in Section \ref{sec:cosmicbirefringence}.

\section{$TB$, or not $TB$? Evidence for a globally nonzero Galactic $TB$ signal}\label{sec:globallynonzero}

We examine the $TB$ signal over the high Galactic latitude sky, defined by the mask described in Section \ref{sec:skymasks}. The {Planck} $TB$ analysis is based on the $353$\,GHz data for both total intensity and polarization \citep{PlanckCollaborationXI:2020}. While $353$\,GHz is the {Planck} frequency channel most sensitive to dust emission in polarization, it is less sensitive than the 545 and 857\,GHz channels to dust total intensity. We can thus compute the cross-power spectra between a total intensity map at one frequency and the polarization maps at another, e.g., $T_{857} B_{353}$.
We compute these cross-correlations from the \texttt{NPIPE} data splits described in Section \ref{sec:data}. Whenever applicable, we compute the estimator for $T_{\nu_1} B_{\nu_2}$ as
\begin{equation}
    T_{\nu_1} B_{\nu_2} = \frac{1}{2} \left(T_{\nu_1}^A B_{\nu_2}^B + T_{\nu_1}^B B_{\nu_2}^A\right) \,,
\end{equation}
where the $A$ and $B$ superscripts denote the two data splits (here assumed to have similar noise properties). We compute the analogous estimator for other quantities. We analyze $D_\ell = \ell(\ell + 1)C_\ell/(2\pi)$, where $C_\ell$ is the pseudo-$C_\ell$ estimator for purified $E$ and $B$ modes \citep{Smith:2006} computed with \texttt{Namaster} \citep{Alonso:2019}. All results presented here are qualitatively insensitive to the choice of $E$ and $B$ purification. We estimate $C_\ell$ in bins of width $\Delta\ell = 19$. The error bars shown in Figure \ref{fig:TB_353_857_HI} represent Gaussian variance only, including contributions from both signal and noise. We also compute the correlation ratio 
\begin{equation}
    r^{XY}_\ell \equiv \frac{C_\ell^{XY}}{(C_\ell^{XX} C_\ell^{YY})^{1/2}},
\end{equation}
where $X$ and $Y$ are any of $T$, $E$, or $B$. 

For $100 \lesssim \ell \lesssim 500$ we find a robustly positive $T_{353}B_{353}$ signal over our fiducial sky mask, with $\left<r_\ell^{TB}\right> \sim 0.05$. The $T_{353}E_{353}$ signal is also robustly positive ($\left<r_\ell^{TE}\right> \sim 0.23$ over the same multipole range). 
A spurious $TB$ correlation could arise from the combination of the real $TE$ signal and imperfect {Planck} polarization angle calibration \citep[e.g.,][]{Abitbol:2016}.
We estimate the polarization angle miscalibration required in order for the measured $T_{353}B_{353}$ to be entirely spurious, $TB^{spurious} = \mathrm{sin}(2 \psi^{miscal})T_{353}E_{353}$ \citep[e.g.,][]{Abitbol:2016}. We find that the Planck polarimeter miscalibration would need to be $\psi^{miscal} \sim 5^\circ$, strongly discrepant with the {Planck} polarization angle calibration uncertainty of $0.28^\circ$ (\citet{2016A&A...596A.110P}, derived from pre-launch measurements described in \citet{Rosset:2010}). 

\citet{Weiland:2020} report a significant nonzero $TB$ signal for $T_{857} B_{353}$ and $T_{545} B_{353}$ computed with the PR3 data. We confirm that result using the \texttt{NPIPE} {Planck} maps (Figure \ref{fig:TB_353_857_HI}). 
\citet{Weiland:2020} also measure a nonzero $T_\mathrm{SFD}B_{353}$ correlation, where $T_\mathrm{SFD}$ is a dust intensity map derived from IRAS and COBE data \citep{Schlegel:1998}, and thus independent of {Planck}. This constitutes evidence that the nonzero $TB$ is not an artifact of {Planck} systematics, except for the global polarization angle miscalibration discussed above. \citet{Weiland:2020} further substitute $B_{353}$ for $B$ derived from either Wilkinson Microwave Anisotropy Probe (WMAP) $K$-band data \citep{Page:2007} or a template map described in \citet{Page:2007} that is derived from optical starlight polarization data \citep{Heiles:2000, Berdyugin:2001, Berdyugin:2004, Berdyugin:2002}. These measurements are independent of the Planck polarization angle calibration.
The $K$-band data probe polarized synchrotron emission while the starlight is largely polarized by the same dust grains probed in the far infrared emission; either way, \citet{Weiland:2020} find evidence for a positive $TB$ correlation, and conclude that measurements of $TB > 0$ are real, rather than spurious. 

As an additional test of the robustness of the $TB$ correlation, we measure \THI$B_{353}$, i.e., the cross-correlation between the \HI4PI 21-cm total intensity and the {Planck} 353\,GHz polarization. The \hi data are entirely independent of the microwave data, and thus this calculation cannot be affected by correlated systematics between {Planck} frequency channels. As summarized in Figure \ref{fig:TB_353_857_HI}, we measure a \THI$B_{353}$ correlation that is consistent with positive $TB$, although the error bars are somewhat larger than the {Planck}-only measurements. Our results support the conclusion that nonzero $TB$ is a real property of the Galactic emission.  

\section{The origin of parity violation in dust polarization}\label{sec:origin}

\subsection{Hypothesis: Nonzero $TB$ from magnetically misaligned dust filaments}\label{sec:hypothesis}

We hypothesize that the observed nonzero $TB_{353}$ is generated, at least in part, by filamentary ISM structures that are misaligned with the projected magnetic field in one preferential direction. A filament-induced $TB$ correlation can be generated by a misalignment between the long filament axis and the plane-of-sky magnetic field orientation \citep{Huffenberger:2020}. The direction of misalignment determines the sign of $TB$, and thus if the statistical misalignment of filaments is skewed to one direction, the global signal will reflect that handedness.

Rather than $E_{353}$ and $B_{353}$, we can consider \Ehi and \BHI, derived from the \HI-based Stokes parameter maps. These maps are constructed by measuring the orientation of filamentary \hi structures: one of the underlying assumptions in the \citet{ClarkHensley:2019} paradigm is that linear \hi structure is preferentially parallel to the magnetic field. 
Filamentary ISM density structures that are aligned with the magnetic field generate a positive $TE$ correlation \citep{Zaldarriaga:2001, Huffenberger:2020}. We thus expect \THI\Ehi $> 0$ by construction. We do not have a reason to expect that \THI\Bhi is robustly nonzero, as no coherent misalignment between the filament axis and \thetahi is included in the construction of the \HI-based Stokes maps. This does not necessarily mean that \thetahi is perfectly aligned with the density structures in any of our $T$ tracers: for one thing, the ISM contains a great deal of structure that is not well described as a linear filament with a single orientation. Also, the \HI-based maps are the integration over the line of sight of 3D Stokes parameter maps constructed by quantifying the \hi morphology in narrow \hi velocity channels. 
Because of this line-of-sight integration, the \thetahi in a given pixel is an intensity-weighted average of the orientations of any \hi structures along the line of sight \citep{Clark:2018}. Still, lacking an expectation of \textit{coherent} misalignment between the \hi intensity and the measured \hi orientation, we expect $T$\Bhi$\sim0$. 

\begin{figure}
    \centering
\includegraphics[width=\columnwidth]{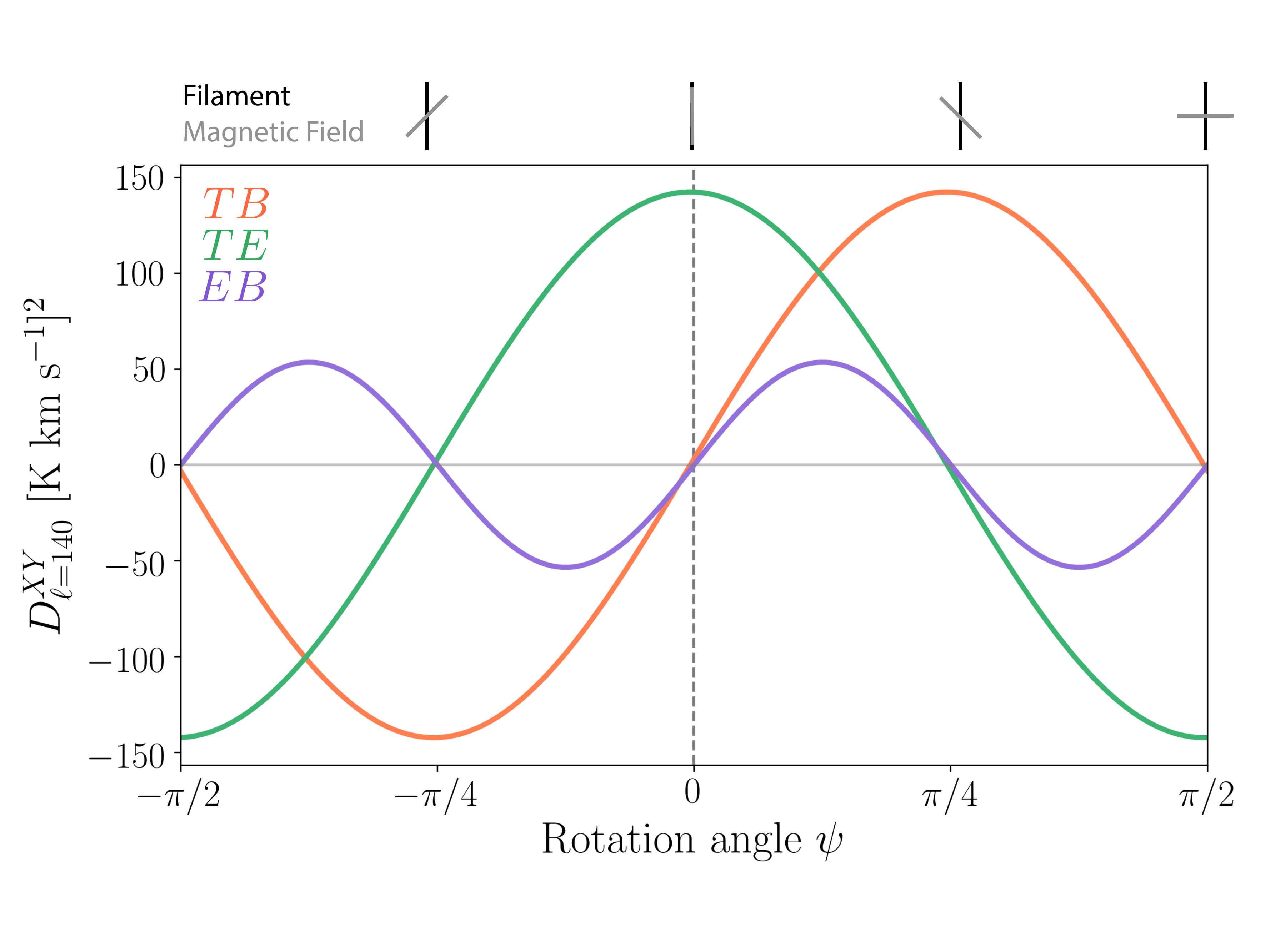}
\caption{\HI$\times$\hi cross-correlations as a function of a uniform rotation angle ($\psi$) applied to $Q_\mathrm{H{\sc I}}$ and $U_\mathrm{H{\sc I}}$, described in Section \ref{sec:hypothesis}. The cross-correlation amplitude is measured at $\ell=140$ for the $TB$ (orange), $TE$ (green), and $EB$ (purple) signals. Top axis illustrates the conceptual meaning of $\psi$. When $\psi=0$, correlations are shown for the raw \citet{ClarkHensley:2019} \HI-based Stokes parameter maps, for which filamentary \hi structures are perfectly aligned with the magnetic field by construction. Nonzero $\psi$ represents a misalignment between \hi structures and the magnetic field, and the $TB$ correlation peaks when $\psi=\pi/4$.   }
\label{fig:uniformrotang}
\end{figure}

We can demonstrate the misaligned filament effect by artificially introducing a global misalignment between \thetahi and the distribution of \hi intensity. To do this, we globally rotate \Qhi and \Uhi by some angle, and cross-correlate the rotated maps with \THI, \QHI, and \UHI. 
We apply this global rotation via

\begin{equation}\label{eq:globalrotmat}
    \begin{bmatrix}
    Q'_\mathrm{H\textsc{i}} \\
    U'_\mathrm{H\textsc{i}}
    \end{bmatrix} =
    \begin{bmatrix}
     \mathrm{cos}(2\psi) & -\mathrm{sin} (2\psi) \\
     \mathrm{sin}(2\psi) & \mathrm{cos} (2\psi)
    \end{bmatrix}
    \begin{bmatrix}
    $\Qhi$ \\
    $\Uhi$
    \end{bmatrix},
\end{equation}
where $\psi$ is the rotation angle applied to each pixel in the \Qhi and \Uhi maps. We then compute \EHI$'$ and \BHI$'$ from \QHI$'$ and \UHI$'$, and show the resulting \HI-based cross-correlations at $\ell=140$ as a function of the rotation angle $\psi$ in Figure \ref{fig:uniformrotang}. The choice of $\ell$ bin does not affect the shape or phase dependence of the signal; the $\ell$ bin affects only the relative amplitudes of $TB$, $TE$, and $EB$. The $\psi=0$ values in Figure \ref{fig:uniformrotang} represent the autocorrelation spectra of the raw (unrotated) \citet{ClarkHensley:2019} maps. These maps display a strong, positive $TE$ signal, and approximately zero $TB$ and $EB$. When $\psi=0$ the $EE/BB$ ratio measured for these maps is also at a maximum. This is consistent with our expectation that these \HI-based maps represent ``perfect alignment" between density structures and the magnetic field by construction. 

The introduction of nonzero $\psi$ represents an artificial, uniform misalignment between the \hi structures and the magnetic field. 
When $\psi=0$, \THI\BHI$'$ is consistent with $0$, and \THI\EHI$'$ is at its maximum value. 
\THI\BHI$'$ is at a maximum when $\psi=\pi/4$ and at a minimum when $\psi=-\pi/4$. This is consistent with our intuitive expectation that a 45$^\circ$ misalignment between a filament and the magnetic field will generate the strongest $B$-mode polarization signal. 

This calculation also clearly demonstrates that if filament misalignment is generating nonzero $TB$, it necessarily also generates nonzero $EB$ (except for $\psi=\pm\pi/4$, but this would yield zero $TE$).  Furthermore, in this simplified misaligned filament model, the sign and magnitude of $EB$ can be predicted by measuring $TE$ and $TB$. This carries important implications that we will return to in Section \ref{sec:cosmicbirefringence}. Here, we will test whether there is evidence for a misaligned filament origin for the nonzero $TB_{353}$ in {Planck} data. 

\begin{figure}
    \centering
\includegraphics[width=\columnwidth]{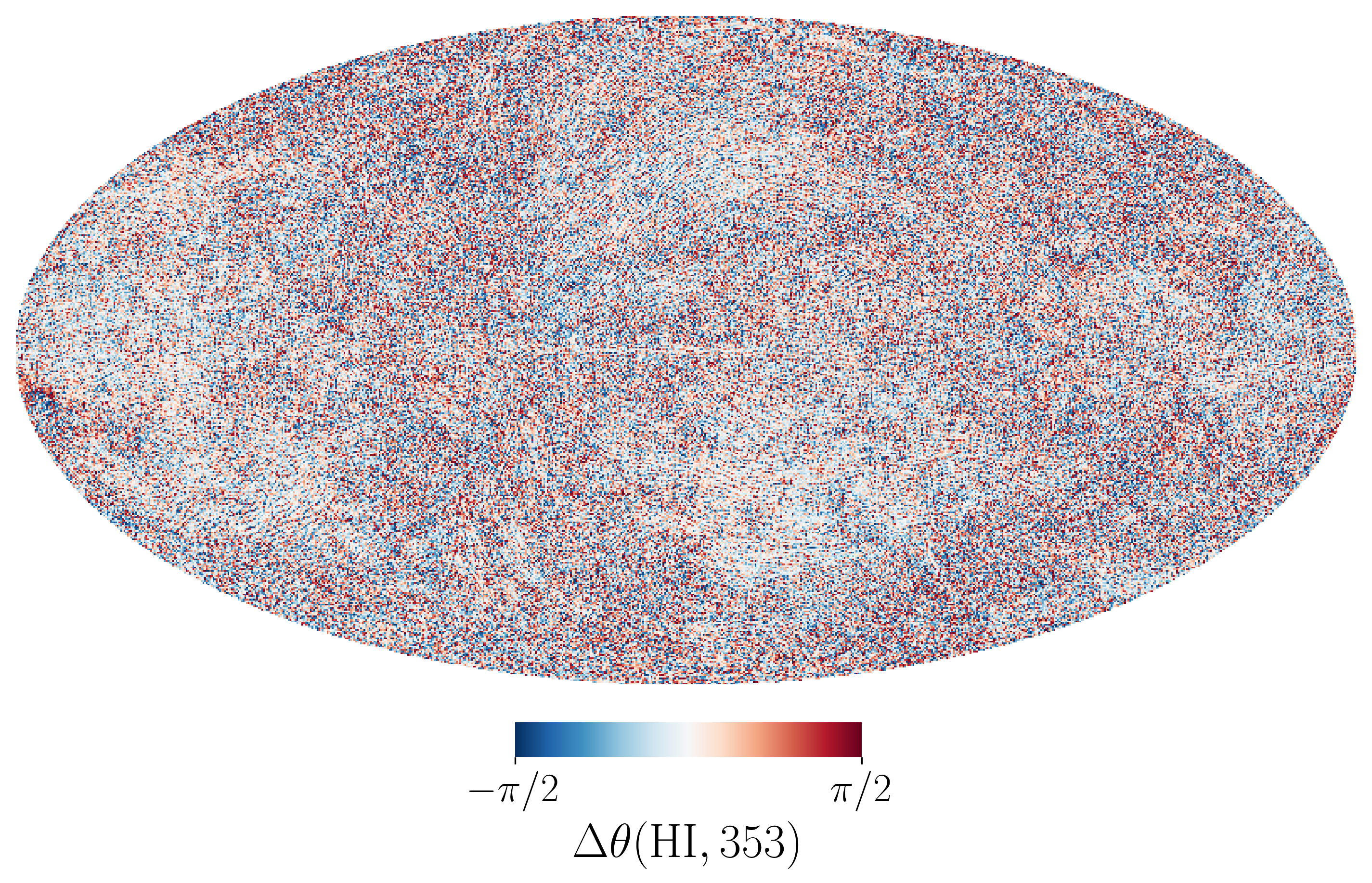}
\caption{Map of \Deltatheta, the signed angular difference between {Planck} $\theta_{353}$ and $\theta_{HI}$ computed from the \citet{ClarkHensley:2019} maps. \DeltathetaS is calculated at $16.2'$ resolution, the native resolution of \HI4PI. This map is in a mollweide projection centered at $(l, b) = (0, 0)$.}
\label{fig:angdiff}
\end{figure}

\subsection{$TB$ is related to \Deltatheta}\label{sec:TBdeltatheta}

We introduce a proxy for the degree of local filament misalignment by quantifying the difference between the 353 GHz polarization angle and the \HI-based polarization angle. 
We define
    
\begin{equation}\label{eq:deltatheta}
    \Delta\theta(1, 2) = \frac{1}{2} \mathrm{arctan}\left( \frac{\sin(2\theta_1)\cos(2\theta_2) - \cos(2\theta_1)\sin(2\theta_2)}{\cos(2\theta_1)\cos(2\theta_2) + \sin(2\theta_1)\sin(2\theta_2)}\right),
\end{equation}
the signed difference between angles $\theta_1$ and $\theta_2$. We apply Equation \ref{eq:deltatheta} to \thetahi and $\theta_{353}$ calculated from the \texttt{NPIPE} full maps to compute \Deltatheta, the signed angular difference between the \HI-based polarization angle and the 353 GHz polarization angle. We compute \DeltathetaS at the $16.2'$ resolution of the \HI4PI data (Figure \ref{fig:angdiff}). 

If the observed nonzero $TB$ is related to a misalignment between ISM density structures and the magnetic field, the observed $TB$ signal should be related to \Deltatheta, our proxy for the angular difference between the orientation of dusty filaments and the local magnetic field. In particular, we expect the sign of \DeltathetaS to be correlated with the sign of $TB$.

\begin{figure}
    \centering
\includegraphics[width=\columnwidth]{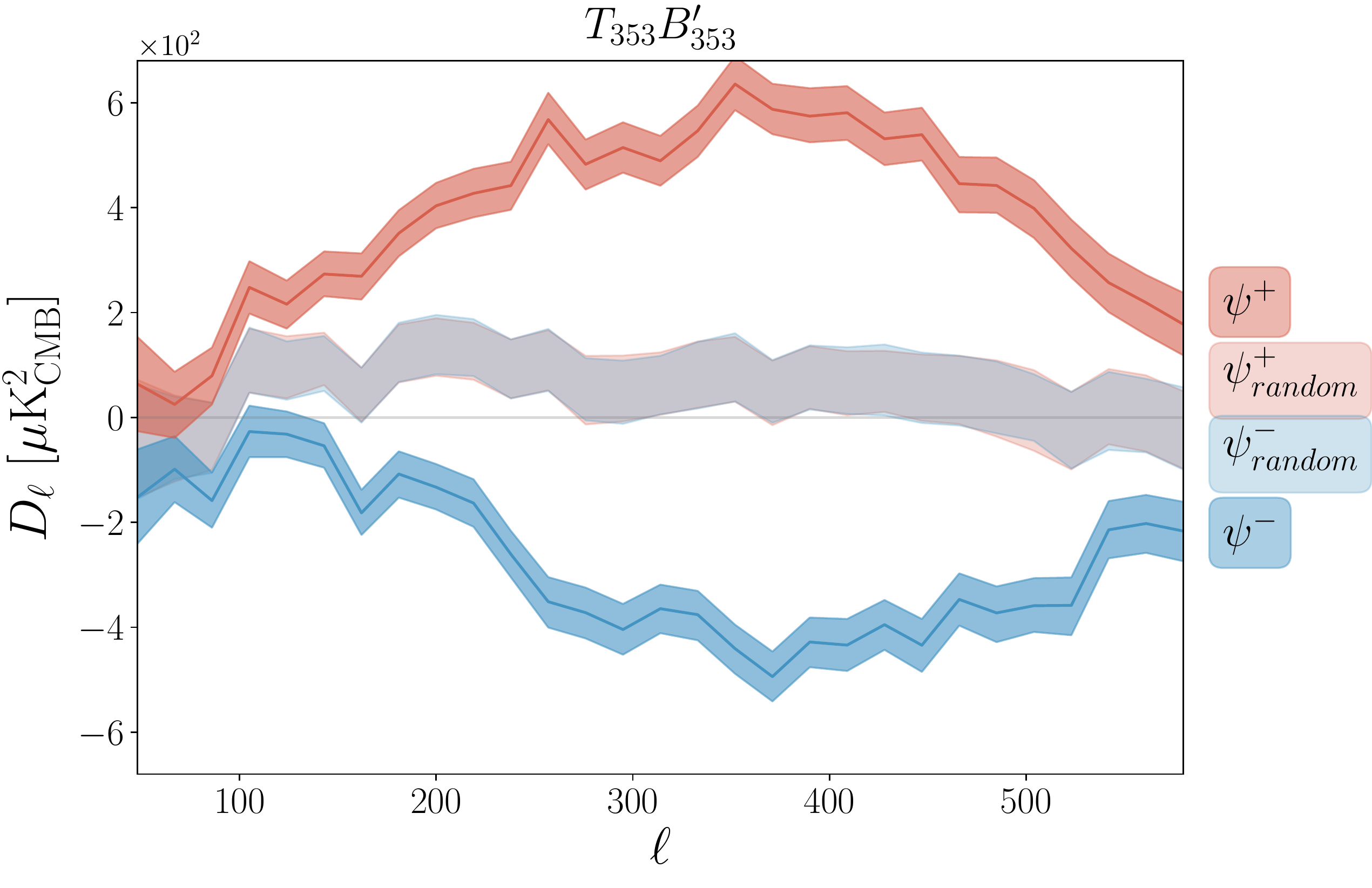}
\caption{$T_{353}B'_{353}$, cross-correlation of the \texttt{NPIPE} 353 GHz total intensity and polarization data for the experiment described in Section \ref{sec:TBdeltatheta}. Red: median (line) and $1\sigma$ spread (contours) of 100 realizations of $T_{353}B_{353}' \pm \sigma_{TB'}$ after rotating the 353 GHz polarization angles of pixels with \DeltathetaS $> 0$ by a random angle ($\psi^{+}$, Equation \ref{eq:psi+}). Blue: analogous calculation with random rotation applied to pixels with \DeltathetaS $< 0$ ($\psi^{-}$, Equation \ref{eq:psi-}).
Lighter contours show null tests (light red: Equation \ref{eq:psirandom+}, light blue: Equation \ref{eq:psirandom-}). $T_{353}B_{353}'$ becomes strongly negative or positive when the 353 GHz polarization angles are randomized based on the sign of \Deltatheta. }
\label{fig:TB_TE_857_353}
\end{figure}

We test a series of modifications to the {Planck} 353 GHz polarization data to test the hypothesis that the Galactic $TB$ signal is related to \Deltatheta. For each test we modify $Q_{353}$ and $U_{353}$ by applying a rotation by an angle $\psi$ to each pixel based on some criterion. We compute 
\begin{equation}\label{eq:rotmat}
    \begin{bmatrix}
    Q_{353}' \\
    U_{353}'
    \end{bmatrix} =
    \begin{bmatrix}
     \mathrm{cos}(2\psi) & -\mathrm{sin} (2\psi) \\
     \mathrm{sin}(2\psi) & \mathrm{cos} (2\psi)
    \end{bmatrix}
    \begin{bmatrix}
    Q_{353} \\
    U_{353}
    \end{bmatrix}
\end{equation}
where $\psi$ is determined based on the sign of \Deltatheta, i.e., $\psi$ is one of

\begin{equation}\label{eq:psi+}
    \psi^{+} = 
\begin{cases}
    \mathcal{R},& \text{if } $\Deltatheta$\,>  0\\
    0,              & \text{otherwise}
\end{cases}
\end{equation}

or

\begin{equation}\label{eq:psi-}
    \psi^{-} = 
\begin{cases}
    \mathcal{R},& \text{if } $\Deltatheta$\,<  0\\
    0,              & \text{otherwise}
\end{cases}
\end{equation}
where $\mathcal{R}$ is a random variable drawn uniformly from the interval $[-\pi/2, \pi/2]$.
The rotation of $Q_{353}$ and $U_{353}$ by a random number preserves the 353 GHz polarized intensity in each pixel, while randomizing the polarization angle. Applying Equation \ref{eq:rotmat} with $\psi = \psi^+$ is equivalent to rotating the 353 GHz polarization angle by a random value in all pixels with positive \Deltatheta, and otherwise leaving the pixels unchanged. 

This formalism allows us to test the influence of the polarization angle structure of pixels with a given sign of \DeltathetaS on the global $T_{353}B_{353}'$ signal. To interpret the results, we also need a null test. We generate 100 map realizations of $\Delta_{syn}$, a Gaussian random field with the same power spectrum as \Deltatheta, and apply Equation \ref{eq:rotmat} to each map with $\psi$ determined by 

\begin{equation}\label{eq:psirandom+}
    \psi^{+}_{random} = 
\begin{cases}
    \mathcal{R},& \text{if } \Delta_{syn}\,>  0\\
    0,              & \text{otherwise}
\end{cases}
\end{equation}

or 

\begin{equation}\label{eq:psirandom-}
    \psi^{-}_{random} = 
\begin{cases}
    \mathcal{R},& \text{if } \Delta_{syn}\,<  0\\
    0,              & \text{otherwise}.
\end{cases}
\end{equation}

For each of these maps we compute $T_{353}B_{353}'$. The results are shown in Figure \ref{fig:TB_TE_857_353}. 
The null tests, i.e., randomizing the 353 GHz polarization angles for pixels selected based on the sign of $\Delta_{syn}$, produce $T_{353}B_{353}'$ correlations that are $\lesssim$ the unrotated $T_{353}B_{353}$: either consistent with 0 or weakly positive. 
By contrast, randomizing the 353 GHz polarization angle for pixels with \Deltatheta$ > 0$ produces a strong positive $T_{353}B_{353}'$ signal. Randomizing the polarization angle of pixels with \Deltatheta$ < 0$ produces a strongly negative $T_{353}B_{353}'$ signal. Evidently, the sign of \Deltatheta\ is predictive of the sign of the {Planck} $TB$ signal.

One interpretation of these experiments is that by randomizing the polarization angle, we ``destroy" the correlations that exist between the total intensity and polarization angle for approximately half of the sky in each test. If there were no correlation between the pixel selection map and the sign of $TB$, this process should only introduce noise to the map, and we should not expect to strengthen the magnitude of $TB$. This is consistent with what we find when we randomize polarization angles based on the sign of $\Delta_{syn}$. Likewise, rotating \textit{all} 353\,GHz polarization angles by $\mathcal{R}$ results in $T_{353}B_{353}' \sim 0$, as expected. By contrast, when we randomize polarization angles based on the sign of \Deltatheta, we seem to preferentially add noise to regions of the sky that give rise to one sign of $TB$. The strong positive $T_{353}B_{353}'$ correlation that we find when we randomize the polarization angles of pixels with \Deltatheta$ > 0$ is consistent with the hypothesis that pixels with \Deltatheta$ > 0$ are preferentially in regions of sky with a negative $T_{353}B_{353}$. Likewise, this suggests that pixels with \DeltathetaS $ < 0$ are preferentially in regions of sky with positive $T_{353}B_{353}$. Note that which sign of \DeltathetaS is associated with a given sign of $TB$ depends on the particular conventions used (n.b. \DeltathetaS = -$\Delta\theta(353,\mathrm{H\textsc{i}})$), but the association of the sign of \DeltathetaS with the sign of $TB$ is robust. 

We consider a number of variations of this experiment and find that they are all consistent with the same hypothesis. We find qualitatively the same result if we replace $\mathcal{R}$ in Equations \ref{eq:psi+} -- \ref{eq:psirandom-} with \Deltatheta, so that rather than rotating by a random angle, we rotate the polarization angles of selected pixels by \Deltatheta. In this variant, $T_{353}B_{353}'$ is nonzero at a higher significance for the maps constructed with $\psi^+$ and $\psi^-$ than for maps constructed with $\psi^+_{random}$ and $\psi^-_{random}$. We also test a different framework: instead of applying Equation \ref{eq:rotmat}, we scramble (resample without replacement) $Q_{353}$ and $U_{353}$ over sets of pixels defined either by the sign of \DeltathetaS or by the sign of $\Delta_{syn}$. This approach changes the sky distribution of the 353 GHz polarized intensity. Nevertheless, we still find that pixel resampling based on the sign of \DeltathetaS strengthens the magnitude of the $T_{353}B_{353}'$ signal more than resampling based on the sign of $\Delta_{syn}$, albeit at lower significance than the rotation-based method that preserves the polarized intensity. 
We find the same behavior for $T_{353}B_{353}'$ and $T_{217}B_{217}'$. If we construct \DeltathetaS from maps of \QHI, \UHI, $Q_{353}$, and $U_{353}$ smoothed to a uniform resolution of FWHM=$30'$, $50'$, or $80'$ and perform the same experiment (without downgrading the pixelization of the maps), we measure the same discrimination between positive and negative $TB$ based on the sign of \Deltatheta, with the largest magnitude of the effect pushed to increasingly lower multipoles as the angular resolution of \DeltathetaS is lowered. 

Could these results be explained by some latent correlation between \DeltathetaS and the 353 GHz polarization angles that is not physically related to a misalignment between \hi structures and the magnetic field? The magnitude of $|$\Deltatheta$|$, for instance, is anti-correlated with the local polarization angle dispersion \citep[see Figure 12 in][]{ClarkHensley:2019}. Physically, this is consistent with the expectation that the dispersion of polarization angles is higher when the mean magnetic field is more parallel to the line of sight \citep[e.g.,][]{Hensley:2019}. This 3D geometry affects the magnitude of \Deltatheta, but does not on its own introduce a preference for the sign of \Deltatheta. 
We find no evidence for a correlation between the sign of \Deltatheta\ and the numerical value of $\theta_{353}$.

The results shown in Figure \ref{fig:TB_TE_857_353} are qualitatively unchanged for \DeltathetaS derived from {Planck} PR3 or \texttt{NPIPE} maps. Deriving \DeltathetaS from alternative \thetahi maps based on the spatial gradient of \HI4PI channel map emission \citep{ClarkHensley:2019} also yields qualitatively similar results. We likewise reproduce the same qualitative results when we calculate $T_{353}B_{353}'$ using the PR3 maps. We conclude that the association between the sign of \DeltathetaS and the sign of $TB$ is not an artifact of any known systematic in the processing of {Planck} data, nor in the calculation of \thetaHI. 

The framework presented here allows us to test hypotheses for the physical nature of $TB$, or any other statistical measure, by randomizing some polarization angles based on test criteria. The conditions on \DeltathetaS in Equations \ref{eq:psi+} and \ref{eq:psi-} enable a test of the hypothesis that the sign of \DeltathetaS is related to the sign of $TB$. The most general expression of our formalism is the application of Equation \ref{eq:rotmat} with $\psi = \psi_{condition}$, where
\begin{equation}\label{eq:psicondition}
    \psi_{condition} = 
\begin{cases}
    \mathcal{R},& \text{if } \mathrm{[condition]}\\
    0,              & \text{otherwise}.
\end{cases}
\end{equation}

We test the additional hypothesis that the sign of $TB$ is related to \nHI, the \hi column density, by applying Equation \ref{eq:psicondition} with conditions on \nHI, e.g., \nhi$ > \mathrm{med}($\nHI) or $\mathrm{P}_{i}($\nHI$) < $ \nhi$ < \mathrm{P}_{i+10}($\nHI), where $P_{i}$ is the $i^{th}$ percentile of \nHI. We find no strong evidence for a correlation between \nhi and the sign of $TB$.

\begin{figure}
    \centering
\includegraphics[width=\columnwidth]{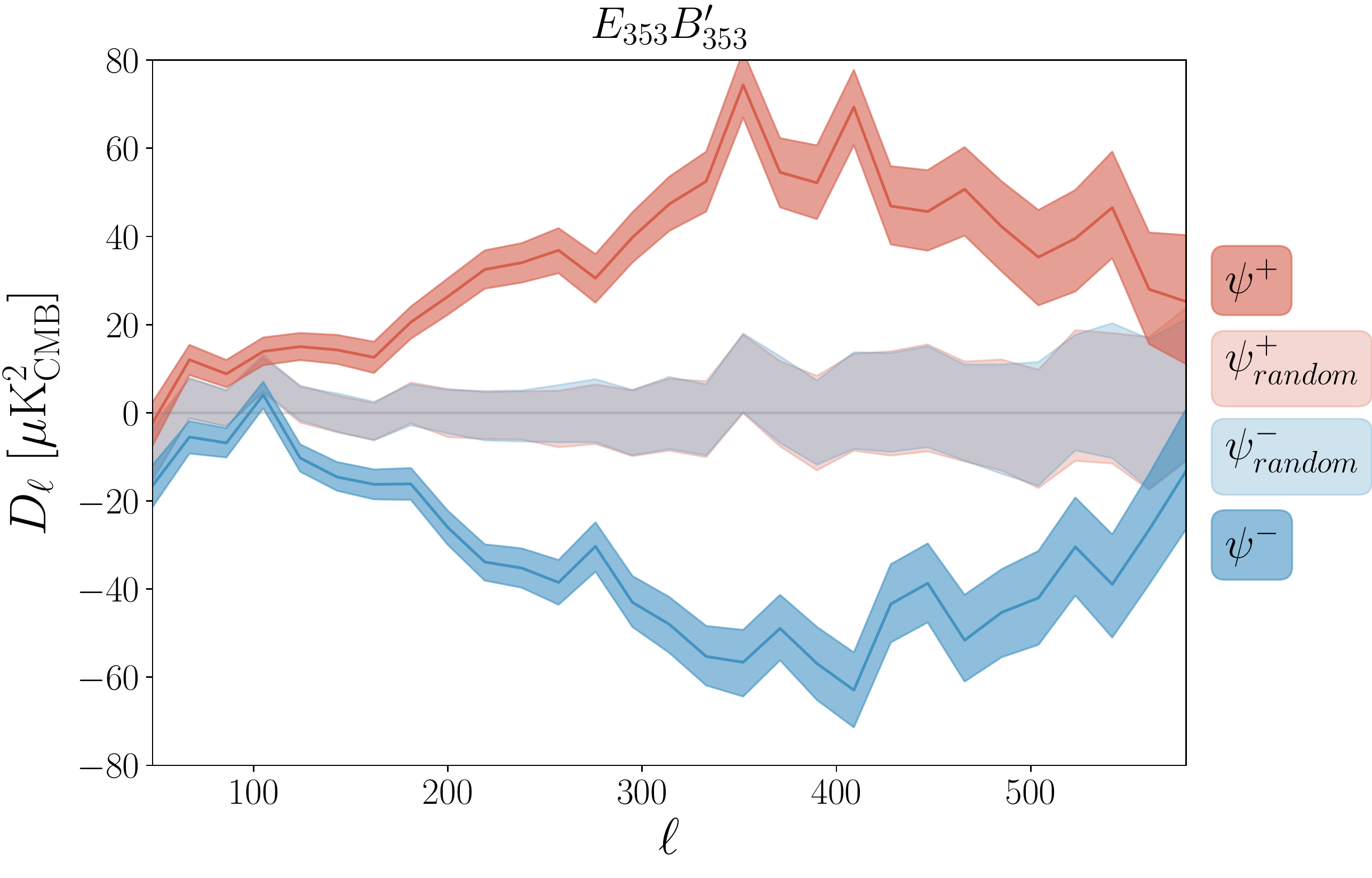}
\caption{The same experiment depicted in Figure \ref{fig:TB_TE_857_353}, but for the $EB$ correlation at 353 GHz (Section \ref{sec:EBdeltatheta}). The cross-correlated maps are the \texttt{NPIPE} A/B splits of the 353 GHz data. Red and blue contours show the $1\sigma$ distribution of 100 realizations of $E_{353}B_{353}' \pm \sigma_{EB'}$ when the polarization angles of the data used to compute $B_{353}'$ are randomized if \DeltathetaS $> 0$ (red) or \DeltathetaS $< 0$ (blue). Light blue and red contours show the $1\sigma$ distribution of the null tests (Equations \ref{eq:psirandom+} and \ref{eq:psirandom-}). }
\label{fig:EBangdiff}
\end{figure}

\subsection{$EB$ is related to \Deltatheta}\label{sec:EBdeltatheta}

If the relative orientation of the magnetic field and dusty filaments is responsible for generating nonzero Galactic $TE$ and $TB$, it follows that these filaments will also generate nonzero $EB$. The relative amplitude of $TB$, $TE$, and $EB$ as a function of misalignment angle is illustrated in Figure \ref{fig:uniformrotang}. The sign of $EB$ is uniquely determined by the combined signs of $TB$ and $TE$ in the misaligned filament paradigm. Because $TE$ is robustly positive over the sky, it follows that $EB$ in this model will have the same sign as $TB$. $TB$, in turn, may change sign depending on the sky mask and angular scale considered, and it follows that the sign of the Galactic $EB$ signal will be mask-dependent as well.

\begin{figure*}
    \centering
    \includegraphics[width=\columnwidth]{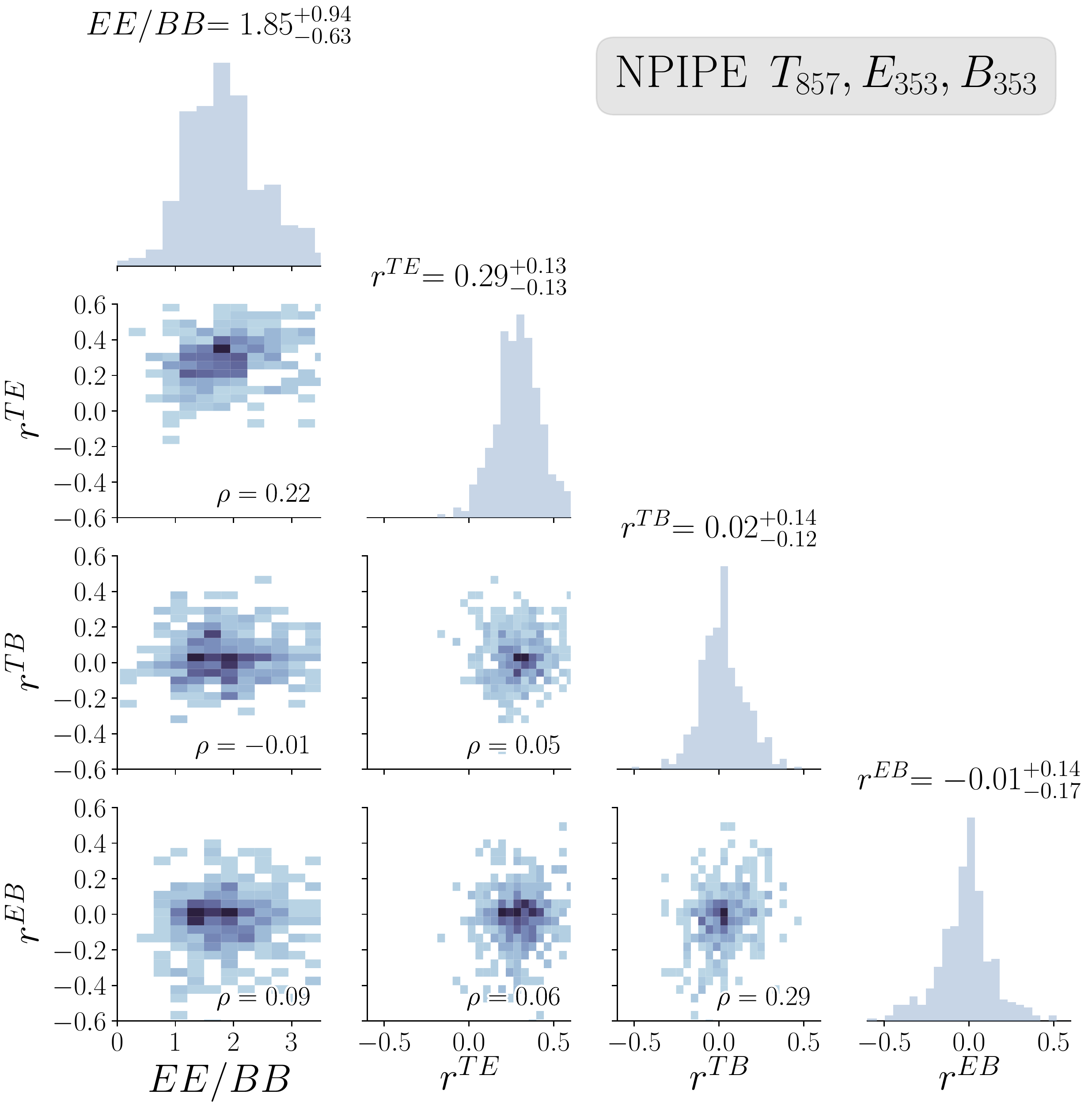}
    \includegraphics[width=\columnwidth]{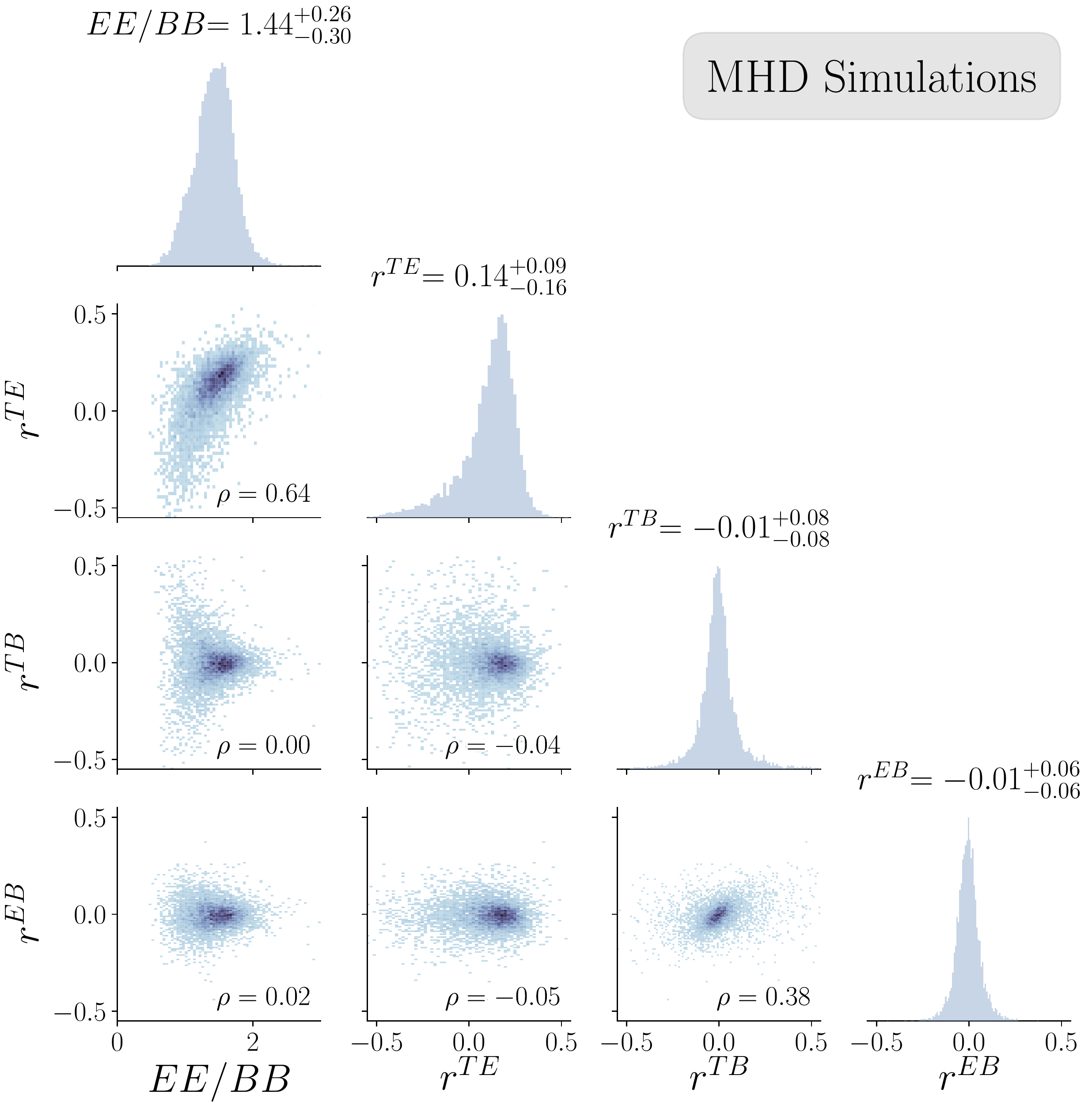}
    \caption{Joint distributions between $EE/BB$ and normalized cross-correlations $r^{TE}$, $r^{TB}$, and $r^{EB}$, computed at $\ell=200$. Left: Correlations between \texttt{NPIPE} $T_{857}$, $E_{353}$, and $B_{353}$ data computed for 12$^\circ$ radius circular regions of data with $|b| > 30^\circ$. 
    Right: Correlations between 3150 synthetic maps from \citet{Kim:2019}. For each map we compute two sets of correlations, for $b > 30^\circ$ and $b < -30^\circ$.
    The Spearman's rank correlation coefficient of each joint distribution is shown in the bottom-right corner of each panel. Diagonal plots show the 1D histograms of each quantity, labeled with the median, $16^{\rm th}$, and $84^{\rm th}$ percentile values.}
    \label{fig:sim_dist}
\end{figure*}

We test whether there is evidence in the {Planck} data for an $EB$ signal associated with \Deltatheta. We apply the same experiment described in Section \ref{sec:TBdeltatheta}, but measure the estimator for the $E_{353}B_{353}'$ cross-correlation between A and B splits of the \texttt{NPIPE} 353 GHz data, where we apply the angle rotation to the data split used to calculate $B_{353}$. The results are shown in Figure \ref{fig:EBangdiff}. We find that randomly rotating the polarization angles of pixels based on the sign of \DeltathetaS yields nonzero $E_{353}B_{353}'$ in excess of associated null tests. As with our $TB$ experiment, applying Equation \ref{eq:psi+} to the 353 GHz data generates positive $E_{353}B_{353}'$ over the $\ell$ range where the \Deltatheta-selected data are distinguishable from the null tests, and applying Equation \ref{eq:psi-} yields negative $E_{353}B_{353}'$. We find the same behavior for $E_{217}B_{217}'$.  We conclude that the sign of $EB$ is related to the sign of \Deltatheta, and that the sign of $EB$ relative to \DeltathetaS has the same sense as the sign of $TB$.   

\subsection{Interpretation: Nonzero $TB$ and $EB$ from magnetically misaligned filaments}\label{sec:interpretation}

The results in Sections \ref{sec:TBdeltatheta} and \ref{sec:EBdeltatheta} indicate that the magnetic misalignment probed by \DeltathetaS is correlated with the $TB$ and $EB$ signals in polarized dust emission. As we predicted, the sign of \DeltathetaS probes the ``handedness" of the local magnetic misalignment of filaments. The globally positive $TB$ signal (Figure \ref{fig:TB_353_857_HI}) thus suggests that there is an overall preference for one handedness of the misalignment over our fiducial sky mask. In our convention, randomizing the polarization angles of pixels with positive \DeltathetaS leads to a positive global $TB$: this suggests that the handedness associated with negative \DeltathetaS is associated with positive $TB$. Put another way, the globally positive $TB$ suggests that there exists a slight preference for ISM density structures within our sky area to be misaligned with $-\pi/2 < $ \DeltathetaS $< 0$ in our convention. It is possible that there is no physical preference for this handedness in the ISM, and that $TB > 0$ is simply the realization of the projected sky that we happen to observe. Alternatively, there may be a true physical preference for this handedness encoded in the formation of dusty filaments. 

\DeltathetaS measures the degree of alignment between the intensity structure traced by HI and the plane-of-sky magnetic field inferred from the 353 GHz dust polarization angle. If the line-of-sight-averaged dust polarization angle traces the integrated, sky-projected magnetic field orientation as is commonly assumed, nonzero \DeltathetaS indicates that the \hi structures are misaligned with the plane-of-sky magnetic field. However, if there exists a coherent misalignment between dust grains and the local magnetic field, the \hi structures could be perfectly aligned with the magnetic field, and we would still measure nonzero \Deltatheta. \DeltathetaS in that case would correspond to the relative orientation of the magnetic field and the alignment direction of interstellar grains. Such discrepancies can arise in the presence of radiative torques from an anisotropic radiation field \citep{Draine:1997}. In the next section, however, we test predictions of our model using MHD simulations that do not model grain alignment and yet agree with our predictions. Thus, grain alignment along a direction other than the local magnetic field is a possible but likely subdominant contributor to the Galactic $TB$ signal.

\subsection{Tests of correlation predictions with data and MHD simulations}\label{sec:correlation}

Sections~\ref{sec:TBdeltatheta} and \ref{sec:EBdeltatheta} demonstrate that \DeltathetaS is predictive of the signs of both $TB$ and $EB$. Because \DeltathetaS varies across the sky (Figure~\ref{fig:angdiff}), this result implies that the global $TB$ and $EB$ signals are mask-dependent quantities. Our results imply that in principle, it is possible to identify regions of sky for which the dust $TB$ and $EB$ are negative. In practice, it may be difficult to define sky masks capable of isolating negative $TB$ in {Planck} data, given the non-trivial spatial structure in \Deltatheta. 

From the magnetically misaligned filament picture we also make predictions for the relative amplitudes of $TB$, $TE$, and $EB$. In particular, if $TE$ is strong and positive ($\psi\sim 0$ in Figure \ref{fig:uniformrotang}), small deviations from perfect alignment between dust filaments and the local magnetic field will generate a correlation between $TB$ and $EB$.
Conversely, if $TE$ is strong and \textit{negative} ($\psi\sim \pm\pi/2$), small deviations from perfect anti-alignment between dust filaments and the local magnetic field will lead to an \textit{anti}-correlation between $TB$ and $EB$. In addition, in the regime where $TB$ is strong and positive ($\psi\sim \pi/4$) or negative ($\psi\sim -\pi/4)$, $TE$ and $EB$ should be correlated or anti-correlated, respectively.

\begin{figure}
    \centering
    \includegraphics[width=\columnwidth]{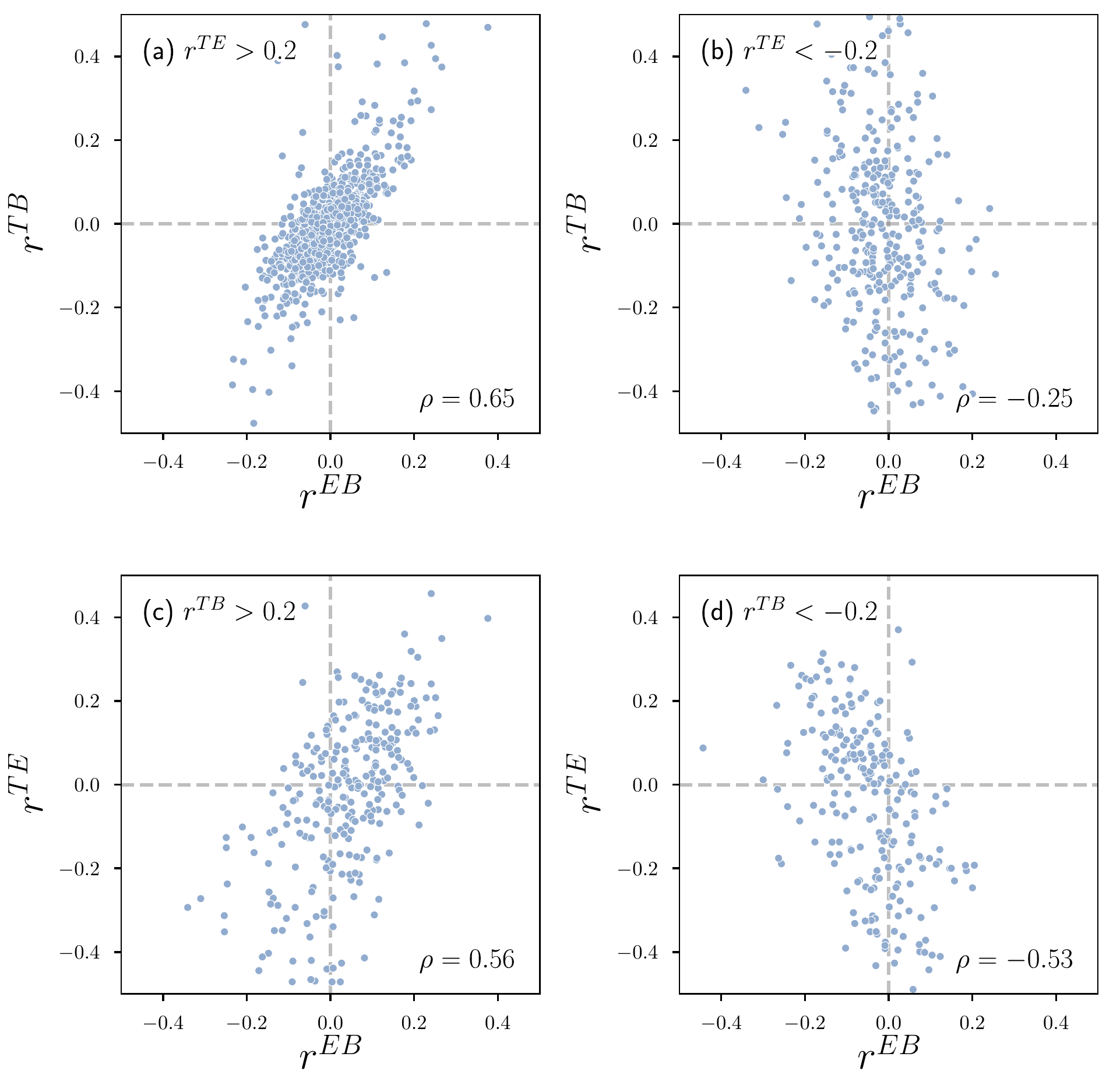}
    \caption{Joint distributions of normalized correlations in selected synthetic maps for (a) strong, positive $TE$ ($r^{TE}>0.2$), (b) strong, negative $TE$ ($r^{TE}<-0.2$), (c) strong, positive $TB$ ($r^{TB}>0.2$), and (d) strong, negative $TB$ ($r^{TB}<-0.2$). Each selection produces the expected correlations as quantified by the Spearman's rank correlation coefficient presented in the bottom-right corner of each panel.}
    \label{fig:sim_test}
\end{figure}

To investigate these correlations in {Planck} data, we tile the sky with $12^\circ$ radius circular regions centered on $N_\mathrm{side}$=8 \texttt{HealPix} pixels \citep{Gorski:2005}, additionally applying our fiducial mask, and apodize each region with a 2$^\circ$ cosine taper \citep[similar to analyses in][]{PlanckCollaborationXXX:2016, Krachmalnicoff:2018, Bracco:2019EB}. We compute cross-correlations between $T_{857}$, $E_{353}$, and $B_{353}$, using splits of the \texttt{NPIPE} data as in the preceding analysis. 
The lefthand panel of Figure \ref{fig:sim_dist} shows the joint distribution between $r^{TE}$, $r^{TB}$, $r^{EB}$, and $EE/BB\equiv C_\ell^{EE}/C_\ell^{BB}$, where $T$ is $T_{857}$ and $E, B$ are $E_{353}, B_{353}$. These values are computed for a multipole bin of width $\Delta\ell=200$ centered at $\ell=200$. The $TE$ correlation is generally strongly positive over the sky regions considered. We find weak positive correlations between $EE/BB$ and $r^{TE}$ (Spearman's rank coefficient $\rho \sim 0.2$) and between $r^{TB}$ and $r^{EB}$ ($\rho \sim 0.3$). The results are similar when we use $T_{353}$ instead. 

\begin{figure*}
    \centering
    \includegraphics[width=\textwidth]{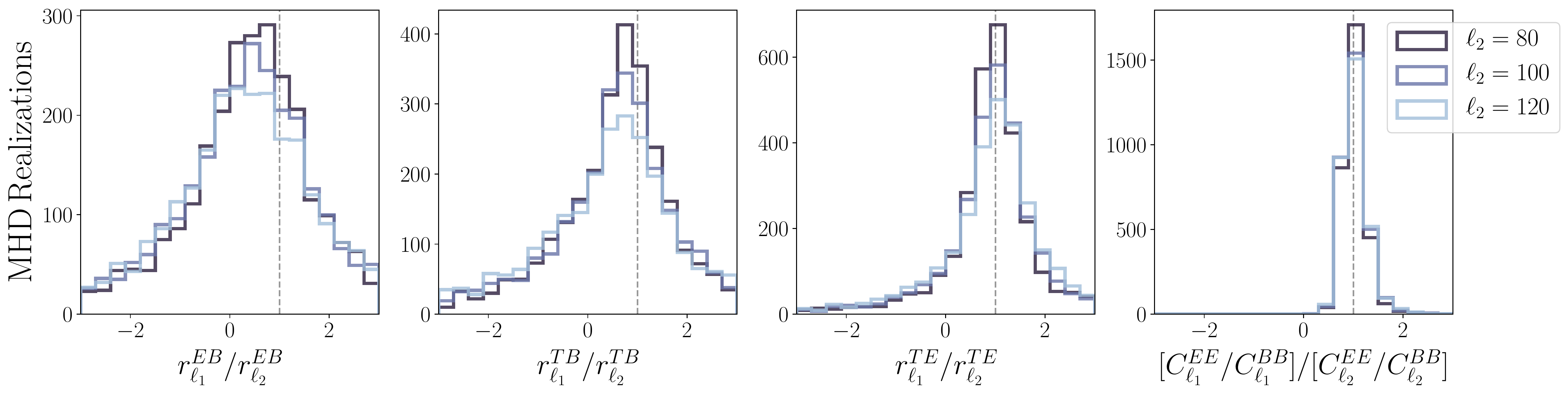}
    \caption{Histograms of the ratio of two $\ell$ bins over 3150 realizations of the MHD simulations. Bin ratios are computed for $r^{EB}$, $r^{TB}$, $r^{TE}$, $EE/BB$, from left to right. Each histogram plots the ratio of $\ell_1=60$ to $\ell_2=80$, $100$, or $120$. }
    \label{fig:sim_hists}
\end{figure*}

The {Planck} data are thus consistent with the behavior we predict when filaments are on average well-aligned with the magnetic field. However, we are limited in our ability to fully test the predicted correlations in data because the sky, even analyzed in small regions, is in the regime where the $TE$ signal is strong and positive ($\psi\sim0$). To further test our predictions, we use synthetic dust polarization maps presented in \citet{Kim:2019}. The maps are derived from MHD simulations based on a solar neighborhood model that includes self-gravity, Galactic differential rotation, cooling and heating, and star formation and supernova feedback \citep{Kim:2017}. A total of 3150 maps with $N_\mathrm{side}$=128 are publicly available.\footnote{\url{https://lambda.gsfc.nasa.gov/simulation/tb_tigress_data.cfm}} For each map we apply a simple hemispheric mask, defined by either $b > 30^\circ$ or $b < -30^\circ$, and compute the pseudo-$C_\ell$ for purified $E$- and $B$-modes using \texttt{Namaster}, as in our analysis of the {Planck} data.
We note that the limited resolution of numerical simulations steepens the power spectra at turbulence dissipation scales (typically $\sim10-20$ pixels). The projection of cubic resolution elements onto a \texttt{Healpix} sky as viewed by an observer placed within the simulation domain gives rise to  non-trivial projection effects, preventing a clear separation of the inertial range and dissipation scales and further steepening the power spectral slopes for the synthetic maps. However, the correlation ratios we measure are robust to these effects, and are converged with different resolutions \citep{Kim:2019}.

The righthand panel of Figure \ref{fig:sim_dist} shows the joint distributions for $r^{XY}$ and $EE/BB$ computed from the simulated maps. The majority of the synthetic maps show $EE/BB>1$ and $r^{TE}>0$, and we find a positive correlation between $EE/BB$ and $r^{TE}$ ($\rho=0.64$). There is also a weak positive correlation between $r^{TB}$ and $r^{EB}$ ($\rho=0.38$). This positive correlation is consistent with the prediction illustrated by Figure \ref{fig:uniformrotang}, as the synthetic maps are dominated by strong, positive $TE$ signals. The $TB$ and $EB$ signals are on average zero. There are no significant correlations among the other quantities. These correlations are qualitatively consistent with the {Planck} data. 

We further test our predictions by selecting realizations of the sky that demonstrate strong $TE$ or $TB$ of a given sign. Figure \ref{fig:sim_test} shows correlations for synthetic maps with (a) strong, positive $TE$ ($r^{TE}>0.2$), (b) strong, negative $TE$ ($r^{TE}<-0.2$), (c) strong, positive $TB$ ($r^{TB}>0.2$), and (d) strong, negative $TB$ ($r^{TB}<-0.2$). Of our 6300 sets of power spectra (North and South hemispheres for 3150 synthetic maps), we find that these criteria are satisfied for (a) 1629, (b) 315, (c) 282, and (d) 232 maps. We confirm our predictions in all regimes. Selecting maps with strong, positive $TE$ significantly enhances the correlation between $TB$ and $EB$ seen in Figure \ref{fig:sim_dist} ($\rho=0.38\rightarrow0.65$). The maps with strong, negative $TE$ show the predicted anti-correlation between $TE$ and $EB$ ($\rho=-0.25$). Finally, while the full suite of simulations shows no correlation between $r^{TE}$ and $r^{EB}$ ($\rho=-0.05$, Figure \ref{fig:sim_dist}), the predicted correlations appear when we select maps based on their $TB$ correlations. Selecting maps with strong positive $TB$ yields the predicted positive correlation between $EB$ and $TE$ (Figure \ref{fig:sim_test} panel c, $\rho=0.56$); selecting maps with strong negative $TB$ yields a negative correlation between $EB$ and $TE$ (panel d, $\rho=-0.53$). In all cases we show the same $\ell=200$ bin that we use to compute correlations in the {Planck} data. We also compute these correlations for a bin of width $\Delta\ell = 40$ centered at $\ell=80$. The measured correlations are comparable or stronger at lower multipole, and still in agreement with predictions: at $\ell=80$ we find (a) $\rho=0.65$, (b) $\rho=-0.41$, (c) $\rho=0.64$, and (d) $\rho=-0.58$.  

In the polarized filament picture, the $EE/BB$ ratio should have some dependence on how well aligned filaments are with the local magnetic field, irrespective of the handedness of any misalignment. Filaments that are well-aligned either parallel or perpendicular to the magnetic field should generate strong $E$-like polarized emission, and thus tend to have higher $EE/BB$ ratios. Filaments with a misalignment angle $\psi \sim \pm \pi/4$ should have strong $B$-like emission, and therefore lower $EE/BB$. We therefore expect that in general, $EE/BB$ will be positively correlated with $\left| r^{TE}\right|$, the absolute magnitude of the $TE$ correlation ratio. Likewise, $EE/BB$ should be negatively correlated with $\left| r^{TB}\right|$. Both of these expectations are borne out in the MHD simulations ($\rho = 0.38, \rho = -0.44$, respectively, for the realizations shown in Figure \ref{fig:sim_dist}). However, the $EE/BB$ ratio is also sensitive to effects other than the degree of magnetic alignment. 
The strength of $EE/BB$ also depends on the aspect ratios of polarized dust filaments: filaments that are longer relative to their widths will tend to have higher $EE/BB$ \citep{Rotti:2019}. Furthermore, polarized emission in both the real sky and the MHD simulations does not solely originate from filamentary structures. It may be the case that filaments contribute a larger proportion of the $TE$, $TB$, and $EB$ amplitudes than the $EE$ and $BB$ amplitudes. 

We investigate the scale dependence of $r_\ell^{EB}$, $r_\ell^{TB}$, $r_\ell^{TE}$, and $EE/BB$ in the simulations by computing ratios of each of these quantities between two $\ell$ bins (Figure \ref{fig:sim_hists}). For this calculation we use the same 3150 maps, but compute cross-correlations over a single $|b| > 30^\circ$ mask for each map, using a fixed bin width of $\Delta\ell = 10$. Figure \ref{fig:sim_hists} shows the ratio of $\ell_1=60$ to $\ell_2=80$, $100$, or $120$ for each of these quantities. We find that the $EE/BB$ ratio is nearly scale-independent over this multipole range, with a histogram sharply peaked at $[EE_{\ell_1}/BB_{\ell_1}]/[EE_{\ell_2}/BB_{\ell_2}] = 1$. By contrast, we find that the synthetic $r^{TE}$, $r^{TB}$, and $r^{EB}$ data are less constant in $\ell$, although $r^{TE}_{\ell_1}/r^{TE}_{\ell_2}$ in particular is still strongly peaked at unity. {Planck} constraints do not find strong evidence for an $\ell$-dependent $TE$ correlation around these angular scales \citep{PlanckCollaborationXI:2020}. As discussed above, the correlation ratios considered here are generally robust to resolution and projection effects. The $\ell$ bins used for the calculation shown in Figure \ref{fig:sim_hists} represent the range over which the resolution convergence is demonstrated in \citet{Kim:2019}.

\begin{figure*}
    \centering
    \includegraphics[width=\textwidth]{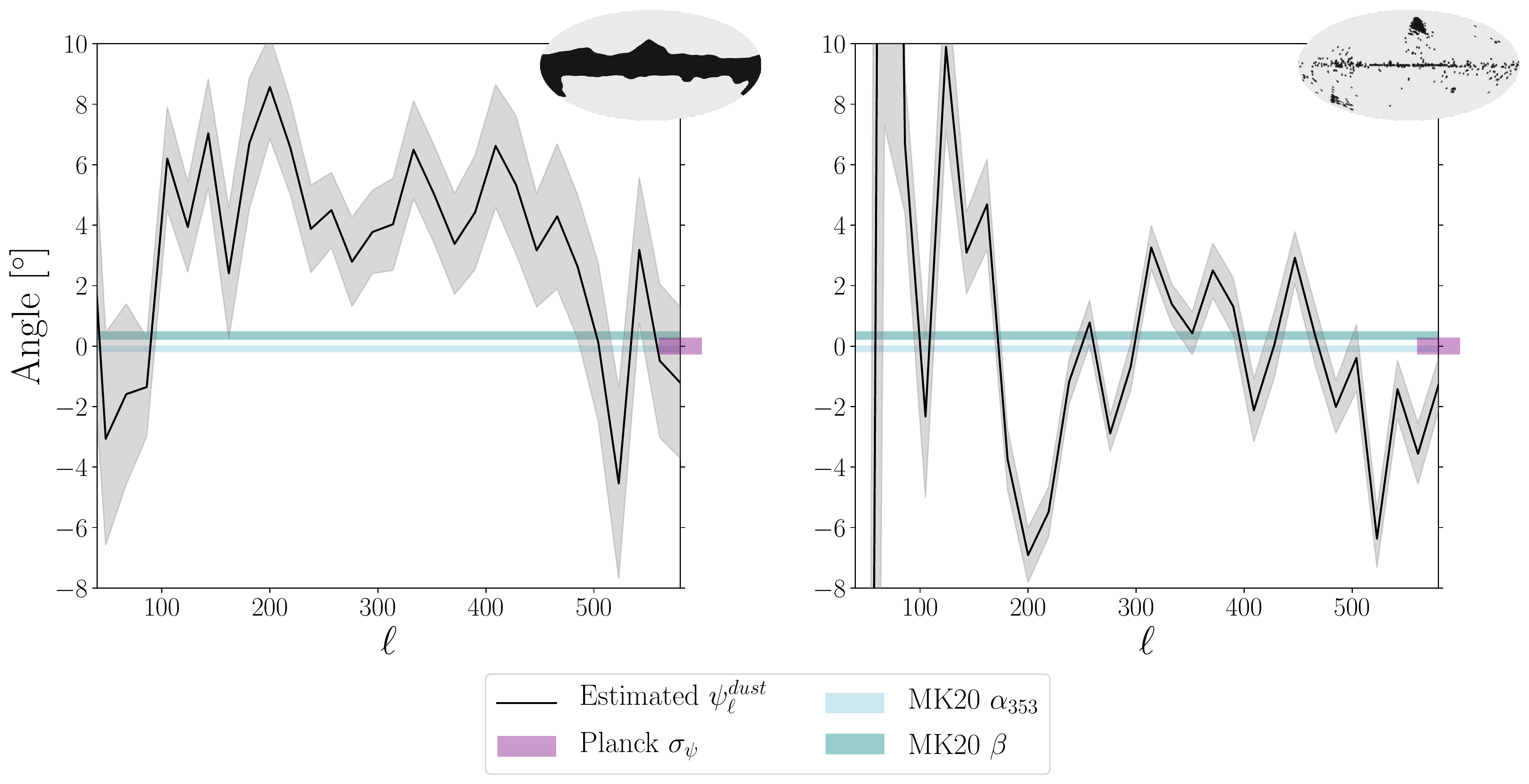}
    \caption{A comparison of our prediction for $\psi^{dust}_\ell$, the effective magnetic misalignment angle, to values from the \citetalias{Minami:2020a} analysis. Black line is the effective $\psi_\ell^{dust}$ calculated from the ratio of $TB$ and $TE$ at 353 GHz. Gray band indicates the propagated $1\sigma$ error. Left: $\psi_\ell^{dust}$ computed over the mask considered in this work ($f_{sky}\sim0.69$, upper righthand corner). Right: the same analysis calculated with PR3 data over the 353 GHz masks used in MK20 (HM1 mask, with $f_{sky}\sim0.92$, pictured in upper righthand corner). Light blue band indicates the MK20 value calculated for the polarization angle miscalibration at 353 GHz, $\alpha_{353} = -0.09^\circ \pm 0.11^\circ$. Teal band denotes the MK20 inference of the isotropic cosmic birefringence angle, $\beta = −0.35^\circ \pm 0.14^\circ$. Purple band indicates the {Planck} polarization angle calibration uncertainty, $\sigma_{\psi} = 0.28^\circ$.    }
    \label{fig:anglecompare}
\end{figure*}

In principle, it may be possible to identify small regions of the sky with negative $TE$, where these correlations could be further tested in real data. Observations of polarized dust emission show evidence that the relative orientation of density structures with respect to the plane-of-sky magnetic field changes from preferentially parallel to more perpendicular at higher column densities \citep{PlanckCollaborationXXXII:2016, PlanckCollaborationXXXV:2016, Soler:2017, Jow:2018}. A magnetic field oriented orthogonally to the main axis of a filament would produce polarized emission with a negative $TE$ correlation \citep{Zaldarriaga:2001, Huffenberger:2020}. It is thus reasonable to expect that high column density regions of sky might have negative $TE$ \citep{Bracco:2019EB}.
This could be tested in measurements of dust emission at higher angular resolution, with $\sim$arcminute-resolution dust polarization maps made by ground-based CMB experiments like the Atacama Cosmology Telescope \citep{Aiola:2020, Naess:2020} and the South Pole Telescope \citep{2014SPIE.9153E..1PB}, as well as next-generation or proposed experiments like the Simons Observatory \citep{Ade:2019}, CMB-S4 \citep{Abazajian:2016,2019arXiv190704473A}, CCAT-prime \citep{2020JLTP..199.1089C}, and PICO \citep{2019arXiv190210541H}.

\section{Implications for cosmic birefringence analyses}\label{sec:cosmicbirefringence}

\citetalias{Minami:2020a} found evidence for a nonzero isotropic cosmic birefringence angle at $2.4\sigma$ significance using {Planck} PR3 data, assuming that the foreground Galactic $EB$ signal vanishes. Our results imply that the Galactic $EB$ signal is generally nonzero, and that the sign of $EB$ can be predicted from the signs of $TE$ and $TB$ measured using the same sky mask. For our fiducial mask, we measure robustly positive $TE$ and $TB$ in {Planck} \texttt{NPIPE} data over $100 \lesssim \ell \lesssim 500$, and thus expect a positive $EB$ contribution from dust. We caution that a different sky mask can yield a different result.

In addition to the sign of the Galactic $EB$ signal, we can estimate its amplitude. Given a measurement of $TB$ and $TE$ in the polarized dust emission, we estimate the global magnetic misalignment angle consistent with these measurements as 

\beq\label{eq:psidust}
\psi^{dust}_{\ell} = \frac{1}{2} \mathrm{arctan} \frac{C_\ell^{TB}}{C_\ell^{TE}}.
\eeq

If we were able to isolate the $TE$ and $TB$ emission from a single filament on the sky, the form of this equation would parameterize the angle between the filament long axis and the local magnetic field orientation (i.e., the quantity for which \DeltathetaS is our proxy). This form also parameterizes a global magnetic misalignment angle, as illustrated in Figure \ref{fig:uniformrotang}. By computing $\psi^{dust}_{\ell}$ as a single $\ell$-dependent quantity over our sky mask, we are parameterizing a scale-dependent ``effective magnetic misalignment" over the sky area considered. This measurement represents the net misalignment angle from the contributions of many dusty filaments, not only the ``one-filament term" (considered by \citet{Huffenberger:2020}, in analogy to the one-halo term in galaxy formation theory). We estimate $\psi^{dust}_{\ell}$ using the \texttt{NPIPE} data splits over our fiducial mask and find $\psi^{dust}_{\ell} \sim 5^\circ$ for $100 \lesssim \ell \lesssim 500$ (Figure \ref{fig:anglecompare}, lefthand panel). 

From $\psi^{dust}_{\ell}$, the predicted sign of the dust $C_\ell^{EB}$ is immediately apparent (e.g., Figure \ref{fig:uniformrotang}). We could estimate the amplitude of the dust $EB$ by treating $\psi^{dust}_{\ell}$ as a global miscalibration angle, such that the amplitude of $EB$ would be proportional to $EE - BB$ \citep[e.g.,][]{Abitbol:2016, Minami:2019}. However, we expect that this treatment will generally overestimate the dust $EB$, because the observed $EE$ and $BB$ contain signal from both filamentary structures, which should contribute substantially to $EB$, and the rest of the dust, which we do not expect to contribute strongly to $EB$. We instead adopt $r_\ell^{EB} \leq r_\ell^{TB}$, and estimate 

\beq\label{eq:rellEBdust}
r_{\ell}^{EB, dust} = r_\ell^{TB} \mathrm{sin} (4\psi_\ell^{dust})
\eeq
as an upper limit on the expected $EB$ correlation ratio.

From Equation~\ref{eq:rellEBdust} and our measurement of $r_\ell^{TB}$ at 353 GHz, we estimate $\left<r_\ell^{EB}\right> \sim 0.017$, where the average is computed on the binned $r_\ell^{EB}$ over $100 \lesssim \ell \lesssim 500$. Considering our measurements of $EE$ and $BB$, this translates to an amplitude $\left<D_\ell^{EB}\right> \lesssim 2.5~\mu {\rm K}_{\rm CMB}^2$ at 353 GHz. This is of the same order as the statistical uncertainty on $EB$ in {Planck} data, although measurements of $EB$ in the 353 GHz data over our mask are largely consistent with our (signed) upper limit prediction. Measuring this intrinsic dust $EB$ should be a target of future microwave polarization experiments. If we further posit that the dust $r_\ell^{EB}$ is constant as a function of frequency, these values can be straightforwardly scaled with the dust SED to estimate the dust $EB$ at any frequency. 

To assess the implications of our work for the \citetalias{Minami:2020a} result, we repeat our analysis over their sky masks. The \citetalias{Minami:2020a} masks are constructed to exclude bad pixels in the PR3 maps and sightlines with bright CO emission. The masks have $f_{sky} \sim 0.92, 0.89$ for half mission (HM) splits 1 and 2, respectively: considerably less conservative than the $f_{sky}\sim 0.69$ mask used in our analysis, which excludes a larger fraction of the bright dust emission near the Galactic plane. The \citetalias{Minami:2020a} masks are tailored for use with the {Planck} PR3 HM1 and HM2 data splits, so we repeat our analysis with those data, as well as with \texttt{NPIPE} A and B data splits. The choice of {Planck} data product does not change the computed $TB$ power spectrum within the errors.
 
We find that $TB$ is neither robustly nonzero nor uniformly positive over the \citetalias{Minami:2020a} sky masks. Thus, we do not predict a uniform sign for $\psi_\ell^{dust}$, nor $C_\ell^{EB}$, over the $\ell$ range considered here. Figure \ref{fig:anglecompare} shows $\psi_\ell^{dust}$ for the \citetalias{Minami:2020a} masks (righthand panel). For comparison, we plot the \citetalias{Minami:2020a} inference of the isotropic cosmic birefringence angle ($\beta$) and simultaneously determined polarization miscalibration angle at 353 GHz ($\alpha_{353}$). These quantities are of the same order, or smaller than, the effective magnetic misalignment angle.

The \citetalias{Minami:2020a} analysis assumes that the intrinsic (physical) foreground $EB = 0$ in their likelihood analysis. Thus the \citetalias{Minami:2020a} method effectively constrains $\beta - \gamma$, where $\beta$ is the birefringence angle and $\gamma$ parameterizes the intrinsic dust $EB$, with positive $\gamma$ corresponding to positive $EB$. If we had inferred that the dust contribution to $EB$ was positive over the \citetalias{Minami:2020a} sky masks, our results would indicate that their measurement of $\beta$ cannot be entirely due to dust, and that the inference of the significance of $\beta$ must be a lower limit. There are two reasons that we cannot draw this conclusion for the \citetalias{Minami:2020a} result. The most important is that the measured $TB$ and inferred dust $EB$ are not robustly positive over the \citetalias{Minami:2020a} masks. Figure \ref{fig:anglecompare} demonstrates that the sign of $\psi_\ell^{dust}$ is not uniform over the range of scales that \citetalias{Minami:2020a} use to infer $\beta$ (in their case, $51 < \ell < 1500$).  

The second reason is that the parameterization of the Galactic $EB$ signal as $C_\ell^{EB, dust} = \frac{1}{2}\mathrm{sin}(4\gamma(\nu)) (C_\ell^{EE, dust} - C_\ell^{BB, dust})$, where $\gamma(\nu)$ is an effective rotation angle, relies on the assumption that the correlation ratio $r_\ell^{EB}$ is constant as a function of $\ell$ \citep{Minami:2019}. Is this assumption well-motivated in the misaligned filaments picture? On the one hand, it is reasonable to expect that magnetically misaligned filaments might be more or less prevalent at particular spatial scales in the ISM. On the other hand, preferred scales in the distribution of filaments could also introduce a scale dependence in quantities like $EE/BB$, and this has not been observed within observational constraints.
It could be that scale dependence exists in the $E$ and $B$ emission from filamentary structures, but this is washed out in the observed $EE/BB$ by the emission from the rest of the ISM, i.e., the diffuse dust not in filaments. 

Empirically, we find that over our fiducial sky mask, $r_\ell^{TB}$ and our inferred $r_\ell^{EB}$ are fairly constant over $100 \lesssim \ell \lesssim 500$. This is not the case, however, for the \citetalias{Minami:2020a} sky area. 
If $r_\ell^{EB}$ has a measurable scale dependence, the foreground $EB$ signal is no longer degenerate with a global polarization angle miscalibration in the \citetalias{Minami:2020a} formalism. The scale dependence of the Galactic $EB$, in addition to its frequency dependence, may then be useful for disentangling this foreground signal from an isotropic cosmic birefringence angle. In this case the foreground $EB$ contribution, whether estimated from the dust $TB$ and $TE$ via our formalism or otherwise, should be included explicitly in the likelihood analysis. As we have shown that the intrinsic Galactic $EB$ is non-negligible, one may expect that the likelihood results will change upon making this correction; the exact shift cannot be predicted without a detailed treatment.

The comparison in Figure \ref{fig:anglecompare} highlights the importance of the sky mask for interpreting the foreground contribution to measurements of cosmic birefringence. Based on our findings, one well-motivated strategy for searches for an isotropic cosmic birefringence angle is to build masks that restrict the Galactic emission to a single sign of $TE$ and $TB$, in order to simplify the expectation for the foreground $EB$. Because $TE$ is generally positive in the diffuse ISM, this approach motivates the exclusion of high-column density sightlines that carry a reasonable astrophysical expectation of a negative $TE$ signal \citep[e.g.,][]{Bracco:2019EB}. Otherwise, the anticorrelation between $TB$ and $EB$ in regions where $TE < 0$ will complicate estimation of the global $EB$ signal that relies on $TB$ and $TE$.

Our findings motivate application of the \citetalias{Minami:2020a} method to the fiducial mask considered here, or similar sky masks where our method predicts a uniformly-signed Galactic $EB$ signal. Our results can also be used to estimate the foreground $C_\ell^{EB}$ directly over a given mask. Further investigation of the evidence for cosmic birefringence using data from current and upcoming CMB experiments will be of great interest.

\section{Conclusions}\label{sec:conclusions}

This paper demonstrates that the observed nonzero $TB$ correlation is related to a misalignment between Galactic dust filaments and the plane-of-sky magnetic field. We summarize our key findings below.

\begin{enumerate}
    \item In agreement with previous analyses, we report a positive $TB$ signal over the high Galactic latitude sky. We measure $TB > 0$ when $B$ is derived from {Planck} \texttt{NPIPE} 353 GHz data and $T$ is any of \texttt{NPIPE} $I_{857}$, $I_{353}$, or a \HI4PI map of \hi column density (Figure \ref{fig:TB_353_857_HI}).
    
    \item We hypothesize that the origin of nonzero $TB$ in Galactic dust emission is a coherent misalignment between ISM dust filaments and the local magnetic field. We use the \citet{ClarkHensley:2019} \HI-based Stokes parameter maps, which predict the dust polarization angle based on the assumption that linear \hi structures are aligned parallel to the local magnetic field. We rotate the \Qhi and \Uhi maps by a fixed angle and cross-correlate the rotated maps with \hi total intensity and with the unrotated \Qhi and \Uhi maps. This exercise demonstrates that nonzero $TB$ and $EB$ can be generated from misalignments between filaments and the magnetic field (Figure \ref{fig:uniformrotang}).
    
    \item We compute \Deltatheta, the angular difference between \thetahi and the {Planck} 353 GHz maps (Figure~\ref{fig:angdiff}). We hypothesize that the sign of \DeltathetaS is predictive of the signs of $TB$ and $EB$ in Galactic dust emission.
    
    \item We introduce a formalism to test our hypothesis. We compute cross-correlations of the {Planck} polarization data after rotating the $Q_{353}$, $U_{353}$ values of about half of the pixels in the map by a random angle. This preserves the polarized intensity but destroys correlations associated with the polarization angles of the selected pixels. As a null test, we select the rotated pixels randomly, and find that the measured $TB'$ signal of the rotated sky is $\lesssim$ $TB$ of the unrotated sky, as expected. If the rotated pixels are instead selected based on the sign of \Deltatheta, we find $TB' > TB$ with $TB'$ strongly positive when we rotate pixels with positive \Deltatheta, and $TB' < TB$ with $TB'$ strongly negative when we rotate pixels with negative \DeltathetaS (Figure \ref{fig:TB_TE_857_353}). This confirms our hypothesis.
    
    \item We further demonstrate that the sign of \DeltathetaS is predictive of the sign of $EB$ in Galactic dust (Figure \ref{fig:EBangdiff}).
    
    \item We predict correlations for the relative amplitudes of $TB$, $TE$, and $EB$, and test these correlations in {Planck} data and in the \citet{Kim:2019} synthetic dust polarization maps of MHD simulations (Figure \ref{fig:sim_dist}). We find strong support for our predictions, particularly in the synthetic data, where we can isolate realizations of the sky with strong positive (negative) $TE$ and observe the predicted positive (negative) correlation between $EB$ and $TB$ (Figure \ref{fig:sim_test}). 
    
    \item Our results strongly support magnetically misaligned ISM filaments as the physical origin of parity-odd signals in Galactic dust emission. Filaments are misaligned relative to the sky-projected magnetic field with either handedness, but a net preference over the sky for one sense of misalignment generates the observed net positive $TB$. 
    
    \item Our results can be used to predict the intrinsic dust $EB$ signal, a critical quantity for searches for an isotropic cosmic birefringence angle. We use the {Planck}-measured $TB$ and $TE$ correlations at 353 GHz to parameterize a scale-dependent effective magnetic misalignment angle, $\psi_\ell^{dust} \sim 5^\circ$ for $100 \lesssim \ell \lesssim 500$ over our fiducial sky mask (Figure \ref{fig:anglecompare}, lefthand panel). Taking the measured $r_\ell^{TB}$ at 353 GHz as an upper limit on $r_\ell^{EB, dust}$, this translates to an estimated intrinsic dust $EB$ of $\left<D_\ell^{EB}\right> \lesssim 2.5~\mu {\rm K}_{\rm CMB}^2$ at 353 GHz. {Planck} data are not sensitive enough to test this prediction, but the intrinsic dust $EB$ is a good target for future experiments. 
    
    \item The intrinsic dust $EB$ is highly mask-dependent. We repeat our analysis over the \citet{Minami:2020a} sky mask, and find that neither the measured $TB$ nor our predicted $EB$ are constant in sign over the $\ell$ range considered (Figure \ref{fig:anglecompare}, righthand panel). \citetalias{Minami:2020a} assume that the intrinsic dust $EB=0$ in their primary analysis, but argue that a positive dust $EB$ would increase the significance of their $2.4\sigma$ inference of a nonzero isotropic cosmic birefringence angle. Our results preclude this interpretation of their measurement as a lower limit because our inference of $\psi_\ell^{dust}$ over the \citetalias{Minami:2020a} mask is not robustly positive, and because we do not find support for the implicit assumption that $r_\ell^{EB} \sim $ constant, a necessary condition for the intrinsic dust $EB$ to be degenerate with the isotropic cosmic birefringence angle.     
    
    \item Based on these findings, we suggest that future searches for cosmic birefringence should include the intrinsic dust $EB$ directly in the model used for the likelihood analysis. Our results can be used to predict or constrain this foreground $EB$. Our results also motivate a careful choice of sky mask in order to simplify the interpretation. 
    
\end{enumerate}

Whether the sign of the preferred filament misalignment is an accident of our particular vantage point on the Galaxy, or reflects some parity violating physics of the ISM, remains an open question. This should be further explored both theoretically and observationally. The work presented here underscores the utility of \hi observations in general, and the \citet{ClarkHensley:2019} maps in particular, for deciphering the physical origin of signals in the Galactic polarized dust emission.

\software{astropy \citep{AstropyCollaboration:2013, AstropyCollaboration:2018}, Healpix \citep{Gorski:2005}, healpy \citep{Zonca:2019}, matplotlib \citep{Hunter:2007}, NaMaster \citep{Alonso:2019}, numpy \citep{Oliphant:2015:GN:2886196}, pandas \citep{mckinney-proc-scipy-2010}}

\section*{Acknowledgments}
We thank David Weinberg and Matias Zaldarriaga for insightful comments on earlier versions of this work. We thank Yuto Minami and Eiichiro Komatsu for sharing the sky masks used in their analysis. 
S.E.C. acknowledges support by the Friends of the Institute for Advanced Study Membership.
C.-G.K. and B.S.H. acknowledge support from the NASA TCAN grant No. NNH17ZDA001N-TCAN.  J.C.H. thanks the Simons Foundation for support. 
This work makes use of observations obtained with {Planck} (http://www.esa.int/Planck), an ESA science mission with instruments and contributions directly funded by ESA Member States, NASA, and Canada.
\HI4PI is based on observations with the 100-m telescope
of the MPIfR (Max-Planck- Institut f\"ur Radioastronomie) at Effelsberg and the Parkes Radio Telescope, which is part of the Australia Telescope and is funded by the Commonwealth of Australia for operation as a National Facility managed by CSIRO. This research has made use of NASA's Astrophysics Data System.

\bibliography{planckrefs,myrefs,references}

\end{document}